\documentclass[a4paper,fleqn,usenatbib]{mnras}

\usepackage{newtxtext,newtxmath}

\usepackage[T1]{fontenc}
\usepackage{ae,aecompl}

\usepackage{graphicx}	
\usepackage{amsmath}	
\usepackage{amssymb}	
\usepackage{siunitx}
\usepackage{arydshln}
\usepackage{multirow}
\usepackage{booktabs}
\usepackage{soul,color}
\usepackage{xcolor}

\title[The buried AGN in NGC 1068]{Probing the circumnuclear absorbing medium of the buried AGN in NGC 1068 through \textit{NuSTAR} observations}

\author[A. Zaino et al.]{A. Zaino,$^{1}$\thanks{E-mail: alessandra.zaino@uniroma3.it (AZ)}
S. Bianchi,$^{1}$
A. Marinucci,$^{2}$
G. Matt,$^{1}$
F. E. Bauer,$^{3,4,5}$
\newauthor W. N. Brandt,$^{6,7,8}$
P. Gandhi,$^{9}$
M. Guainazzi,$^{10}$
K. Iwasawa,$^{11}$
S. Puccetti,$^{2}$
\newauthor C. Ricci,$^{12,13}$
D. J. Walton$^{14}$
\\
$^{1}$Dipartimento di Matematica e Fisica, Università degli Studi Roma Tre, via della Vasca Navale 84, I-00146 Roma, Italy \\
$^{2}$Agenzia Spaziale Italiana (ASI) -- Unità di Ricerca Scientifica, Via del Politecnico snc, I-00133 Roma, Italy \\
$^3$Instituto de Astrof\'isica and Centro de Astroingenier\'ia, Facultad de F\'isica, Pontificia Universidad Cat\'olica de Chile, Casilla 306, \\ Santiago 22, Chile \\
$^{4}$Millennium Institute of Astrophysics (MAS), Nuncio Monse\~nor S\'otero Sanz 100, Providencia, Santiago, Chile \\
$^{5}$Space Science Institute, 4750 Walnut Street, Suite 205, Boulder, Colorado 80301, USA \\
$^{6}$Department of Astronomy and Astrophysics, The Pennsylvania State University, 525 Davey Laboratory, University Park, PA 16802, \\ USA \\ 
$^{7}$Institute for Gravitation and the Cosmos, The Pennsylvania State University, University Park, PA 16802, USA \\ 
$^{8}$Department of Physics, 104 Davey Laboratory, The Pennsylvania State University, University Park, PA 16802, USA \\
$^{9}$Department of Physics and Astronomy, University of Southampton, Highfield, Southampton, SO17 1BJ, UK \\
$^{10}$European Space Agency (ESA) -- European Space Research and Technology Centre (ESTEC), Keplerlaan 1, NL-2201 AZ Noordwijk, \\ The Netherlands \\ 
$^{11}$ICREA and Institut de Ciències del Cosmos, Universitat de Barcelona, IEEC-UB, Mart\'i i Franquès, 1, E-08028 Barcelona, Spain \\
$^{12}$N\'ucleo de Astronom\'ia de la Facultad de Ingenier\'ia, Universidad Diego Portales, Av. Ej\'ercito Libertador 441, Santiago, Chile \\
$^{13}$Kavli Institute for Astronomy and Astrophysics, Peking University, Beijing 100871, China \\
$^{14}$Institute of Astronomy, University of Cambridge, Madingley Road, Cambridge CB3 0HA, UK \\
}

\date{Accepted 2020 January 10. Received 2020 January 9; in original form 2019 October 7.}

\pubyear{2020}

\begin{document}
\label{firstpage}
\pagerange{\pageref{firstpage}--\pageref{lastpage}}
\maketitle

\begin{abstract}
We present the results of the latest \textit{NuSTAR} monitoring campaign of the Compton-thick Seyfert 2 galaxy NGC 1068, composed of four $\sim$50 ks observations performed between July 2017 and February 2018 to search for flux and spectral variability on timescales from 1 to 6 months.
We detect one unveiling and one eclipsing event with timescales less than 27 and 91 days, respectively, ascribed to Compton-thick material with $ N_H=(1.8\pm0.8)\times10^{24}$ cm$^{-2}$ and $N_H\geq(2.4\pm0.5)\times10^{24}$ cm$^{-2}$ moving across our line of sight.
This gas is likely located in the innermost part of the torus or even further inward, thus providing further evidence of the clumpy structure of the circumnuclear matter in this source.
Taking advantage of simultaneous \textit{Swift}-XRT observations, we also detected a new flaring ULX, at a distance  $d\sim30"$ (i.e. $\sim$2 kpc) from the nuclear region of NGC 1068, with a peak X-ray intrinsic luminosity of $(3.0\pm0.4)\times10^{40}$ erg s$^{-1}$ in the 2-10 keV band.

\end{abstract}




\section{Introduction}

According to a widely accepted paradigm, AGN are powered by accretion of matter onto supermassive black holes (SMBHs) located in galactic centers \citep{Lynden-Bell1969}.
It is now well known that most AGN are \lq\lq obscured\rq\rq\ in the X-rays (e.g. \citealp{Bianchi2012} and references therein; \citealp{RamosAlmeidaRicci2017}) and at least part of their observed X-ray variability is likely due to the variations of the circumnuclear medium surrounding the central engine (e.g. \citealp{Risaliti2002}; \citealp{Yang2016}).
As stated by one popular AGN unification model (\citealp{Antonucci1993}; \citealp{Netzer2015}), this obscuring medium is optically thick, composed of dust and gas and arranged in an axisymmetric dusty structure with luminosity dependent dimensions of 0.1-10 pc (i.e. the \lq\lq torus\rq\rq).
Its column density is large enough to completely obscure the central source in some directions.
The inclination of the line of sight with respect to the dusty torus is able to explain broadly the observational differences between Type 1 and Type 2 AGN, being the former observed at low angles with respect to the torus axis, while the latter are observed at high inclinations, with the nuclear region completely hidden by the torus itself.
However, although X-ray studies have confirmed, to first order, the widely accepted unified scheme, the location, geometry and physical state of the absorbing material are still widely debated \citep{Bianchi2012}.
In particular, the observation of short timescale variations ($\sim$days or even $\sim$hours) in the absorption column density in several nearby bright sources, such as NGC 1365 (\citealp{Risaliti2005}; \citealp{Rivers2015a}), NGC 4151 \citep{Puccetti2007}, NGC 4388 \citep{Elvis2004} and NGC 7582 (\citealp{Bianchi2009}; \citealp{Rivers2015b}), implies that the obscuring medium has a clumpy geometry, consisting of a series of discrete clouds likely at sub-pc scales.
Although the density of this absorbing matter likely increases towards the equatorial plane (e.g. \citealp{Nenkova2008a} and \citealp{Nenkova2008b}), these variations are at odds with the classical torus geometry, which invokes a smooth distribution of dust and gas in a uniform toroidal structure (\citealp{PierKrolik1992}; \citealp{PierKrolik1993}; \citealp{Fritz2006}).

NGC 1068 ($D_L=14.4$ Mpc; \citealp{Tully1988}) is one of the best known Seyfert 2 galaxies partly because the unification model was first proposed to explain the presence of broad optical lines in its polarized light \citep{AntonucciMiller1985}.
According to its classification, the nucleus of NGC 1068 is heavily obscured by dust.
Near- and mid-infrared observations spatially resolved the dust structures within this galaxy, revealing a torus consistent with a two-component dust distribution.
In particular, VLTI/VINCI observations for the first time favored a multi-component model for the torus intensity distribution.
Taken into account also K-band speckle interferometry, these data showed that a part of the flux originates from scales clearly smaller than $\sim$0.4 pc, arising from substructures of the dusty torus or from the central accretion flow viewed through moderate extinction, and another part of the flux from larger scales of the order of $\sim$3 pc (\citealp{Wittkowski2004}, and references therein).
MIDI observations confirmed this scenario; in particular, \citet{Jaffe2004} modelled the observed spectra using two components with different size and temperature, each of which is a two-dimensional Gaussian aligned with the axis parallel to the radio jet.
They considered a central hot component (T>800 K) marginally resolved along the source axis and surrounded by a warm dust component (T$\sim$320 K) in a structure 2.1 parsec thick and 3.4 parsec in diameter.
More recent MIDI observations allowed resolution of the nuclear mid-infrared emission from NGC 1068 in unprecedented detail, with a maximum resolution of 7 mas (i.e. $\sim$0.5 pc at the distance of NGC 1068).
In particular, \citet{Raban2009} found that the mid-infrared emission can be represented by two components, each with an elliptical Gaussian brightness distribution: a compact hot component (T$\sim$800 K), 1.35 parsec long and 0.45 parsec thick in FWHM (i.e. the inner funnel of the obscuring torus), tilted by $\sim$45\textdegree\ with respect to the radio jet and with similar size and orientation to the observed water maser distribution \citep{Gallimore2004}, and a 3$\times$4 pc warm component (T$\sim$300 K) marking the colder and extended part of the torus-like structure.

Recent ALMA observations have been able to resolve the molecular torus in NGC 1068 over spatial scales of $D$=10-30 pc, demonstrating that there is radial density stratification as well as hints of counter-rotation and a high velocity outflow (\citealp{Imanishi2018}; \citealp{Garcia-Burillo2019}; \citealp{Impellizzeri2019}).
The observed physical parameters are significantly different depending on which line transition used to image the torus, highlighting its many faces.
In particular, the CO(2-1), CO(3-2) and HCO$^+$(4-3) maps provided a full-size of the torus $D_{CO(2-1)}=28\pm0.6$ pc, $D_{CO(3-2)}=26\pm0.6$ pc and $D_{HCO^+(4-3)}=11\pm0.6$ pc, respectively \citep{Garcia-Burillo2019}.

From an X-ray point of view, the multi-epoch X-ray spectra of NGC 1068 were analysed by \citet{Bauer2015} using different observatories, including a \textit{NuSTAR} pointing in 2012, and spanning a time period of $\sim$16 years.
The authors modeled the broadband emission of the source with a combination of a completely obscured transmitted power law ($\Gamma=2.10_{-0.07}^{+0.06}$), scattering by both warm and cold reflectors, radiative recombination continuum and line emission, and off-nuclear point-source emission, being the latter due to the ULX population within the NGC 1068 field of view.
In particular, the reflected emission was due to a multi-component reflector with three distinct column densities, in which the higher $N_H$ component ($N_{H,1}\sim10^{25}$ cm$^{-2}$) provides the bulk of the flux of the Compton hump, the lower $N_H$ component ($N_{H,2}=(1.4\pm0.1)\times10^{23}$ cm$^{-2}$) contributes primarily to the iron line emission and reproduces the curvature of the continuum around 10 keV, and a third reflector on more extended scales (>140 pc) provides almost 30\% of the neutral iron K$\alpha$ line flux (see Appendix \ref{model} and \citealp{Bauer2015} for further details).

NGC 1068 was observed again in 2014 and in 2015 with a joint \textit{XMM-Newton} and \textit{NuSTAR} monitoring campaign during which a transient excess above 20 keV was observed and ascribed to a Compton-thick unveiling event in which material with $N_H\geq2.5\times10^{24}$ cm$^{-2}$ moved temporarily out of our line of sight \citep{Marinucci2016}, allowing the intrinsic radiation to pierce through the circumnuclear medium.
VISIR and MIDI observations performed before and immediately after the X-ray variations showed constant behavior in the infrared emission of the nuclear region of NGC 1068, confirming the hypothesis that the observed change in the X-ray regime was not due to an intrinsic change in the luminosity of the central accretion disk, but to escaping emission through the patchy torus clouds \citep{Lopez-Gonzaga2017}.
However, due to the large separation of the \textit{NuSTAR} observations ($\sim$6 months), only an upper limit of $\sim$2 pc could be given on the location of such a variable absorber, which is consistent both with the BLR and the torus.
In the latter case, the change in the absorbing column density along the line of sight required to explain the X-ray variability of the AGN in NGC 1068 could be associated with the small-scale structure of the molecular torus imaged by ALMA \citep{Garcia-Burillo2019}.

\begin{table*}
	\centering
	\small
	\caption{\textit{NuSTAR} observation log for NGC 1068.}
	\label{tab:obsID}
	\begin{tabular}{cccccccc}
		\toprule
			   &	obsID		&	Start	time				&	Stop	time				&	Detector	&	Net exposure	&	\multicolumn{2}{c}{Net count rate $^b$ [counts s$^{-1}$]}	\\
			   &				&						&						&			&	time $^a$ [ks]	&	3-5.5 keV 				&	 20-79 keV			\\
		\midrule
		OBS1 &	60302003002	&	2017-07-31 00:16:09		&	2017-08-01 03:26:09		&	FPMA	&	50.0			&	0.0246$\pm$0.0007		&	0.0289$\pm$0.0009		\\
	   		   &				&						&						&	FPMB	&	49.8			&	0.0240$\pm$0.0007		&	0.0262$\pm$0.0008		\\
		OBS2 &	60302003004	&	2017-08-27 20:51:09		&	2017-08-29 03:36:09		&	FPMA	&	52.5			&	0.0253$\pm$0.0007		&	0.0332$\pm$0.0009		\\
			   &				&						&						&	FPMB	&	52.4			&	0.0256$\pm$0.0007		&	0.0309$\pm$0.0009		\\
		OBS3 &	60302003006	&	2017-11-06 03:31:09		&	2017-11-07 06:31:09		&	FPMA	&	49.7			&	0.0254$\pm$0.0007		&	0.0301$\pm$0.0009		\\
			   &				&						&						&	FPMB	&	49.5			&	0.0236$\pm$0.0007		&	0.0281$\pm$0.0008		\\
		OBS4 &	60302003008	&	2018-02-05 05:26:09		&	2018-02-06 11:36:09		&	FPMA	&	54.6			&	0.0313$\pm$0.0008		&	0.0276$\pm$0.0008		\\
			   &				&						&						&	FPMB	&	54.5			&	0.0299$\pm$0.0008		&	0.0261$\pm$0.0008		\\
		\bottomrule
		\end{tabular}
		\raggedright \textit{Notes.} $^a$ Net exposure time, after screening was applied on the data. $^b$ Net source count rate after screening and background subtraction, as observed in the 3-5.5 keV and 20-79 keV energy ranges.
\end{table*}

In this paper, we present results from the latest \textit{NuSTAR} monitoring campaign of NGC 1068 performed between July 2017 and February 2018 to sample variability timescales from 1 to 6 months.
The aim was to search for flux and spectral variability on timescales not covered during the previous monitoring campaign and providing tighter constraints on the circumnuclear absorbing Compton-thick material, its physical properties, and its distance from the illuminating source.

The paper is organized as follows. In Section \ref{Section2} we discuss the X-ray observations and data reduction, with Sections \ref{NuSTAR_obs} and \ref{Swift_obs} devoted to \textit{NuSTAR} and \textit{Swift}-XRT data, respectively.
In Section \ref{Section3} we describe the X-ray spectral analysis, while in Section \ref{Section4} we discuss our findings, comparing them with the previous X-ray results and, finally, in Section \ref{Section5} we summarize our work.
The detection of a new ULX is discussed in Appendix \ref{ULX}, while a detailed description of the best-fit model adopted by \citet{Bauer2015}, which is the basis of our modelling, is in Appendix \ref{model}.
Finally, Appendix \ref{cal_line} is devoted to calibration uncertainties.

Throughout the paper, we assumed a flat $\Lambda$CDM cosmology with H\textsubscript{0}=70 km s\textsuperscript{-1} Mpc\textsuperscript{-1}, $\Omega$\textsubscript{$\Lambda$}=0.73 and $\Omega$\textsubscript{m}=0.27, i.e. the default ones in XSPEC 12.10.1 \citep{Arnaud1996}.

\section{Observations and data reduction}
\label{Section2}

	\subsection{\textit{NuSTAR}}
	\label{NuSTAR_obs}

NGC 1068 was observed by \textit{NuSTAR} \citep{Harrison2013} with its two co-aligned X-ray telescopes, with corresponding Focal Plane Modules A (FPMA) and B (FPMB), during a monitoring campaign composed of four observations of about 50 ks each, performed between July 2017 and February 2018 (see Table \ref{tab:obsID}).

The Level 1 data products were processed using the \textit{NuSTAR} Data Analysis Software (NuSTARDAS) package (version 1.8.0).
Event files (Level 2 data products) were extracted using the \textit{nupipeline} task, adopting standard filtering criteria and the latest calibration files available in the \textit{NuSTAR} calibration database (CALDB 20180126).
The source spectra were extracted from circular regions of radius 50 arcsec, corresponding to an encircled energy fraction (EEF) of about 70\%, from both FPMA and FPMB.
The same radius was used to extract background spectra, selecting a region on the same chip, uncontaminated by source photons or background sources.
The net exposure times and the total count rates after this process are reported in Table~\ref{tab:obsID} for each spectrum and for both FPMA and FPMB.

Finally, the two \textit{NuSTAR} spectra from each observation were binned in order not to oversample the instrumental energy resolution by a factor larger than 4 and to have a signal-to-noise ratio (SNR) greater than 7 in each background-subtracted spectral channel.
This ensures the applicability of the $\chi^2$ statistic to evaluate the quality of spectral fitting, avoiding a dramatic oversampling due to the high flux of the source.
The same energy bins were used for both FPMA and FPMB.

	\subsection{\textit{Swift}}
	\label{Swift_obs}
In our work we also analyzed two \textit{Swift}-XRT archival observations performed simultaneously with \textit{NuSTAR} in November 2017 and February 2018 and a ToO observation requested in June 2018 (see Table \ref{tab:obsID_sw}).

\begin{table}
	\centering
	\small
	\caption{\textit{Swift}-XRT observation log for NGC 1068.}
	\label{tab:obsID_sw}
	\begin{tabular}{cccc}
		\toprule
		obsID		&	Start	time				&	Exposure time 		&			\\
					&						&		[ks]			&			\\
		\midrule
		0088104005	&	2017-11-06 08:40:57		&		2.0			&	(1)		\\
		0088104006	&	2018-02-05 05:27:57		&		2.1			&	(2)		\\
		0088104007	&	2018-06-15 16:13:48		&		1.6			&	(3)		\\
		\bottomrule
	\end{tabular}
	\raggedright \textit{Notes.} (1) Simultaneous to OBS3. (2) Simultaneous to OBS4. (3) ToO observation.
\end{table}

To analyze the XRT observation taken in February 2018, we extracted a source spectrum from a circular region with a 20 arcsec radius centered on the ULX (see Appendix \ref{ULX} for further details), ensuring that the nuclear emission of NGC 1068 is excluded.
A background spectrum was also extracted from a source-free circular region with a 60 arcsec radius.
We used identical extraction regions for the other two XRT spectra (November 2017 and June 2018), generating spectra with \textsc{Xselect}, effective area files with \textsc{Xrtmkarf}, and using the redistribution matrix file \textsc{swxpc0to12s6$_{-}$20130101v014.rmf}.
Finally, all the XRT spectra were rebinned in order to have at least 15 counts per energy bin.

\section{X-ray spectral analysis}
\label{Section3}

\begin{figure}
	\centering
	\includegraphics[width=1.0\columnwidth]{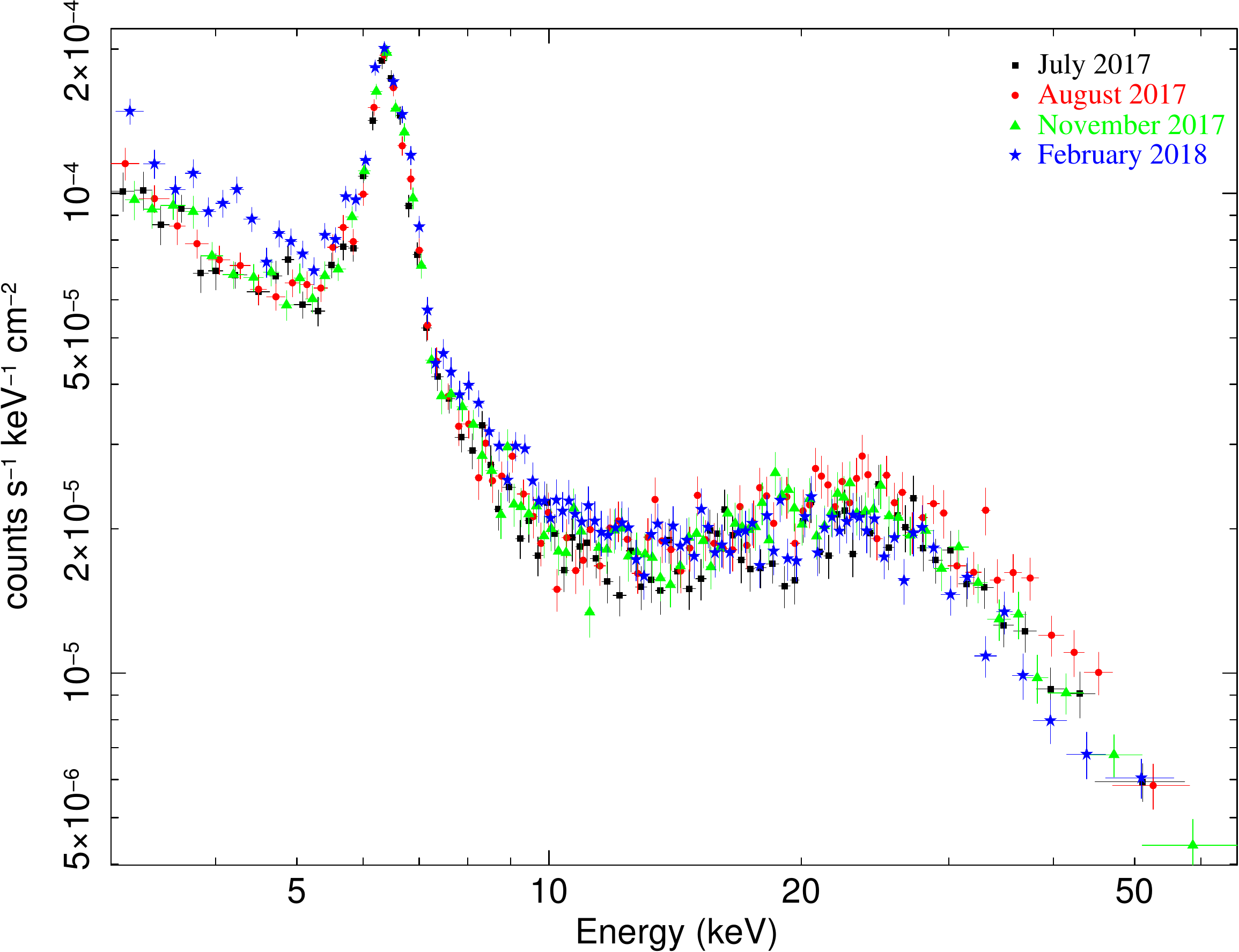}
    \caption{\textit{NuSTAR} monitoring spectra taken in July 2017 (black squares), August 2017 (red circles), November 2017 (green triangles) and February 2018 (blue stars). FPMA and FPMB data of each observation are grouped together for clarity purposes only.}
    \label{fig:spectra}
\end{figure}

The X-ray spectral analysis of NGC 1068 was performed using XSPEC 12.10.1 \citep{Arnaud1996} and the $\chi^2$ statistic, apart from the ULX analysis (see Appendix \ref{ULX}), where the C-statistic \citep{Cash1976} was used.
The photoelectric cross sections for all absorption components are those from \citet{Verner1996}, while the element abundance pattern is from \citet{Wilms2000} and the metal abundance is fixed to solar.
Unless stated otherwise, errors correspond to the 90 per cent confidence level for one interesting parameter ($\Delta \chi^2$=2.706).
The four \textit{NuSTAR} monitoring spectra are shown in Figure \ref{fig:spectra}.

Focusing on the soft X-ray band below 5.5 keV, we can observe an excellent agreement of the data taken in July, August and November 2017 (see black, red and green symbols in Figure \ref{fig:spectra}, respectively).
However, an unexpected increase of the flux by $25\pm2\%$ was observed in February 2018 with respect to the previous observations (see blue circles in Figure \ref{fig:spectra} and the count-rates in Table \ref{tab:obsID}).
\textit{Swift}-XRT observations reported in Table \ref{tab:obsID_sw} confirmed the appearance of a transient source at a distance of $\sim$2 kpc (i.e. $\sim$30 arcsec) from the nucleus of NGC 1068 as the origin of this excess (see Appendix \ref{ULX} for further details).

\begin{figure*}
	\centering
	\includegraphics[width=2.088\columnwidth]{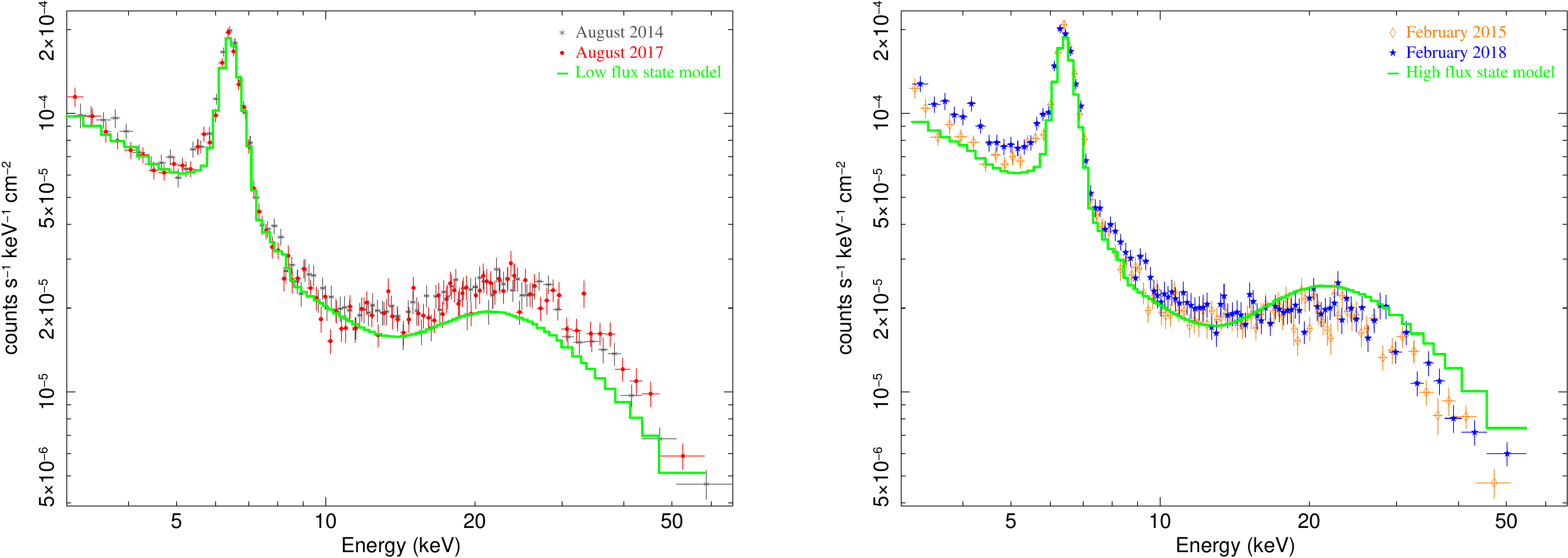}
	\caption{\textit{NuSTAR} observations of NGC 1068 during the last two \textit{NuSTAR} monitoring campaigns. \textit{Left panel.} Comparison between OBSID 60002033002 (August 2014; grey asterisks) and OBS2 (August 2017; red circles), catching the source in a high flux state. \textit{Right panel.} Comparison between OBSID 60002033004 (February 2015; orange diamonds) and OBS4 (February 2018; blue stars), observing the source in a low flux state. In both panels, the green solid line represents the modelling of the opposite flux state, and FPMA and FPMB data of each observation are grouped together for clarity purposes only. The disagreement between blue data and the high flux model below 5.5 keV is due to the appearence of the ULX in February 2018. We refer to the text for further details.}
    \label{fig:spectraHS-LS}
\end{figure*}

	Considering the whole \textit{NuSTAR} band (3-79 keV), we find that the iron line emission does not vary significantly during the whole monitoring (see Figure \ref{fig:spectra}); however, the count-rate above 20 keV clearly varies by up to $\sim$20\% between the observations (see Table \ref{tab:obsID}).
This behavior is similar to that observed three years previously; in particular, NGC 1068 was caught in a higher flux state in August 2014 and 2017 and in a lower one in February 2015 and 2018, as reported in Table \ref{tab:HS-LS}. Additionally, we find excellent overlap of the spectral shapes in the high and low states during the two \textit{NuSTAR} campaigns (see Figure \ref{fig:spectraHS-LS}).
Both these findings suggest that we are observing eclipsing/unveiling events affecting only the spectrum above 10 keV, as found in \cite{Marinucci2016}.
The benefit of the current \textit{NuSTAR} monitoring campaign is that we have two additional observations in July and November 2017 (see Table \ref{tab:obsID}) that enable us to more accurately probe the variability of NGC 1068.

\begin{table}
	\centering
	\small
	\caption{Comparison between the higher and the lower flux states of NGC 1068 during 2014-2015 and 2017-2018 \textit{NuSTAR} monitoring.}
	\label{tab:HS-LS}
	\begin{tabular}{cccc}
		\toprule
		\multicolumn{2}{c}{High state}			&	\multicolumn{2}{c}{Low state}			 \\
		Date			&	Count rate $^{(1)}$	&	Date			&	Count rate $^{(1)}$	\\
		\midrule
								\multicolumn{4}{c}{FPMA}							\\
		Aug 2014		&	$0.032\pm0.001$	&	Feb 2015		&	$0.026\pm0.001$	\\
		Aug 2017		&	$0.033\pm0.001$	&	Feb 2018		&	$0.028\pm0.001$	\\
		\midrule
								\multicolumn{4}{c}{FPMB}							\\
		Aug 2014		&	$0.031\pm0.001$	&	Feb 2015		&	$0.023\pm0.001$	\\
		Aug 2017		&	$0.031\pm0.001$	&	Feb 2018		&	$0.026\pm0.001$	\\
		\bottomrule
	\end{tabular}
	\raggedright \textit{Notes.} $^{(1)}$ Net count rate in the 20-79 keV band, expressed in counts/s.
\end{table}

\begin{figure}
	\centering
	\includegraphics[width=0.98\columnwidth]{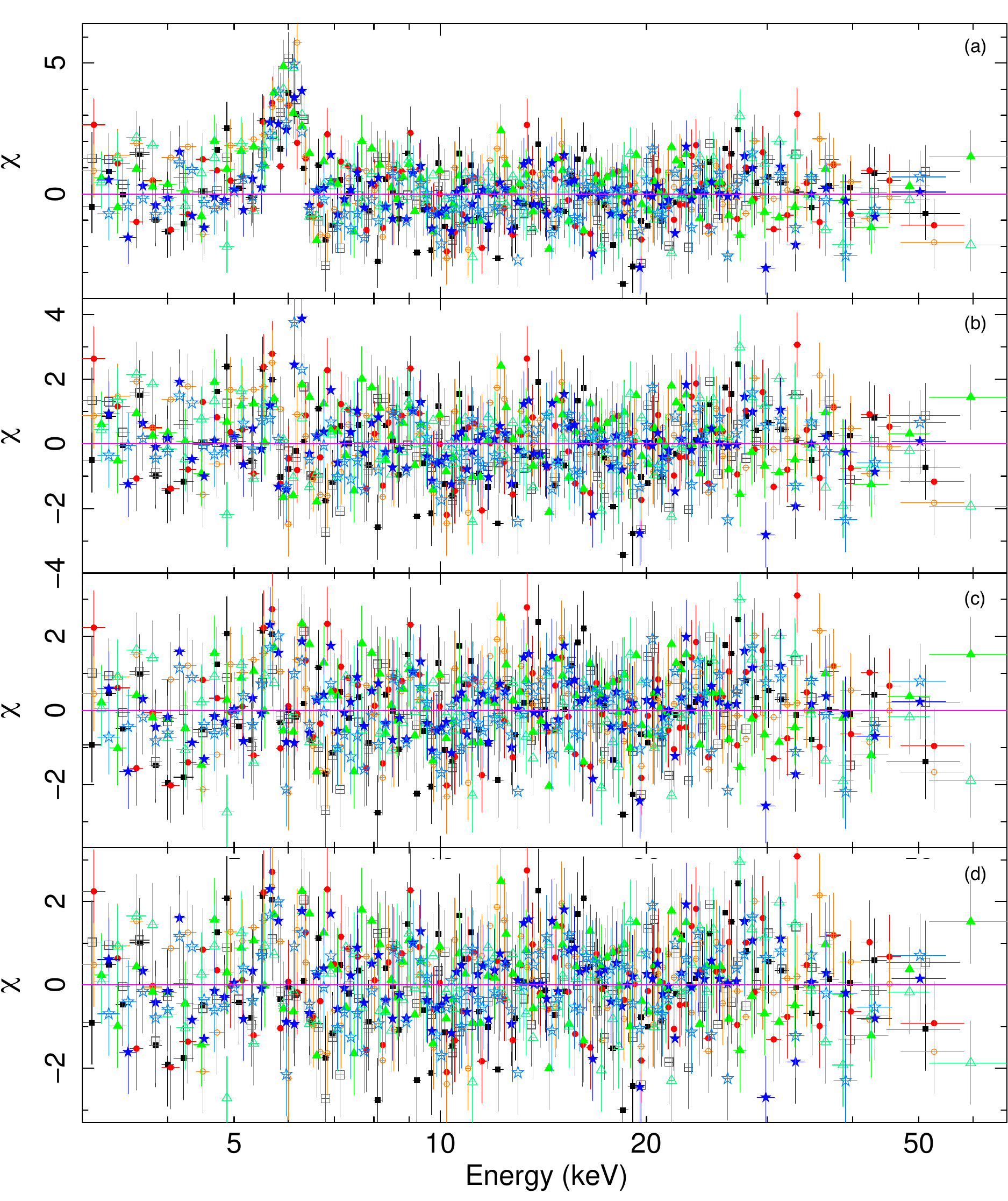}
	\caption{\textit{From top to bottom.} Residuals with respect to Model A, B, C and D (solid magenta line in the four panels) plotted in terms of sigmas. OBS1, OBS2, OBS3 and OBS4 are shown in black and grey squares, red and orange circles, green and dark green triangles and blue and light blue stars, while filled and empty symbols represent FPMA and FPMB data, respectively.}
    \label{fig:fit}
\end{figure}

To model our data set, we adopted the best-fit model discussed in \citealp{Bauer2015} (see Appendix \ref{model} for a detailed description of the model), with an additional cut-off power-law component to account for the ULX contribution in OBS4 (Model A in Table \ref{tab:models}), whose spectral analysis is described in Appendix \ref{ULX}.
Due to the fact that ULXs break to very steep spectra above $\sim$10 keV (e.g. \citealp{Stobbart2006}, \citealp{Gladstone2009}, \citealp{Pintore2017}, \citealp{Walton2018}), we expect that this new source should have a negligible effect on the high-energy data from the nucleus.
In our configuration, the reflected emission is reproduced by three distinct reflectors ($\theta_1=\ang{90}$, $N_{H,1}\geq9.7\times10^{24}$ cm$^{-2}$; $\theta_2=\ang{0}$, $N_{H,2}=(1.4\pm0.1)\times10^{23}$ cm$^{-2}$; $\theta_3=\ang{0}$, $N_{H,3}=(5.0_{-1.9}^{+4.2})\times10^{24}$ cm$^{-2}$) modeled with MYTORUS tables in a decoupled configuration \citep{Yaqoob2012}, where the normalizations for the different angles vary independently, while the continuum and line components of a given angle are fixed, corresponding to a patchy torus distribution.
Applying this model to our data set, the fit was not good ($\chi_r^2=1.63$), mainly due to significant residuals at $\sim$6 keV (see panel (a) in Figure \ref{fig:fit}).

Taking advantage of previous \textit{NuSTAR} observations, some of which were performed simultaneously with \textit{XMM-Newton}, we argue that the significant residuals observed in Model A have no astrophysical origin, but are spurious calibration features (see Appendix \ref{cal_line} for a detailed analysis).
We modeled this component out of the residuals adopting an additional emission line (i.e. a \texttt{gauss} component in XSPEC), whose parameters are reported in Table \ref{tab:models} (Model B), obtaining a significant improvement of the fit ($\chi_r^2=1.18$ -- see panel (b) in Figure \ref{fig:fit}).
We stress that this further component does not affect our conclusions.

\begin{table*}
	\centering
	\small
	\caption{Summary of the six models discussed in the text and differences with respect to the Bauer model summarized in Appendix \ref{model}.}
	\label{tab:models}
	\begin{tabular}{ccccccccccc}
		\toprule
										\multicolumn{11}{c}{\textbf{Model A:} Bauer model + ULX contribution in OBS4}				\\
		\midrule
		$\rm{\Gamma_{ULX}}$	&&&		$\rm{E_c}$ [keV]	&&&			$\rm{norm_{ULX}}$ [ph cm$^{-2}$ s$^{-1}$ keV$^{-1}$]		&&&&		$\rm{\chi^2/dof}$		\\[0.5ex]
		$1.5\pm0.4$			&&&		$\geq$5			&&&		$(2.6_{-0.6}^{+0.7})\times10^{-4}$							&&&&		1186/730				\\[0.5ex]
	\end{tabular}	\\[0.8ex]
	\begin{tabular}{cccc}
		\midrule
										\multicolumn{4}{c}{\textbf{Model B:} Model A + \texttt{gauss} component at $\sim$6 keV}					\\
		\midrule
					&	$\rm{E_{\texttt{gauss}}}$	(FPMA/FPMB)	&	$\rm{norm_{\texttt{gauss}}}$ (FPMA/FPMB)	&		$\rm{\chi^2/dof}$			\\[0.5ex]	
					&			[keV]						&		[ph cm$^{-2}$ s$^{-1}$ keV$^{-1}$]		&								\\[0.5ex]
		OBS1		& 	$6.1\pm0.1$ / $6.1\pm0.1$			&	$1.7\pm0.5$ / $2.5\pm0.5$				&		\multirow{4}{*}{840/714}		\\[0.5ex]	
		OBS2		& 	$6.1\pm0.1$ / $6.1\pm0.1$			&	$1.3\pm0.5$ / $2.3\pm0.5$				&								\\[0.5ex]	
		OBS3		&	 $6.0\pm0.1$ / $6.1\pm0.1$			&	$2.1\pm0.5$ / $2.2\pm0.5$				&								\\[0.5ex]	
		OBS4		&	 $5.9\pm0.1$ / $5.9\pm0.1$			&	$1.7\pm0.4$ / $1.9\pm0.5$				&								\\[0.5ex]	
	\end{tabular}	\\[0.8ex]
	\begin{tabular}{ccccccccccccc}
		\midrule
										\multicolumn{13}{c}{\textbf{Model C:} Model B with reflectors parameters free to vary, but assumed to be the same between the observations}															\\
		\midrule
		$\rm{\Gamma}$	&&&	$\rm{N_{H,1}}$	&	$\rm{N_{H,2}}$	& $\rm{N_{H,3}}$	&&&	$\rm{norm_1}$			&$\rm{norm_2}$					&	$\rm{norm_3}$			&&	$\rm{\chi^2/dof}$	\\[0.5ex]
						&&&	\multicolumn{3}{c}{[$10^{24}$ cm$^{-2}$]}				&&&		\multicolumn{3}{c}{[ph cm$^{-2}$ s$^{-1}$ keV$^{-1}$]}								&&					\\[0.5ex]	
		$2.10\pm0.01$		&&&	$10^{\dag}$	&	$0.13\pm0.01$	&	$\geq4.3$		&&&	$(3.0\pm0.2)\times10^{-1}$&	$(4.3_{-0.4}^{+0.3})\times10^{-2}$	&$(0.9\pm0.3)\times10^{-2}$	&&	784/724			\\[0.5ex]

	\end{tabular}	\\[0.8ex]
	\begin{tabular}{ccccccccccccc}
		\midrule
										\multicolumn{13}{c}{\textbf{Model D:} Model C with only $N_H$ along the l.o.s. and intrinsic flux free to vary}	\\
		\midrule
					&&&&		$\rm{N_{H}^{l.o.s.}}$	[$10^{24}$ cm$^{-2}$]		&&&&&	$\rm{norm_{intr}}$ [ph cm$^{-2}$ s$^{-1}$ keV$^{-1}$]			&&&		$\rm{\chi^2/dof}$			\\[0.5ex]
		OBS1		&&&&			$\geq7.3$				&&&&&	$5.0_{-4.6}^{+2.1}$				&&&		\multirow{4}{*}{784/730}		\\[0.5ex]	
		OBS2		&&&&		$6.3_{-1.0}^{+1.2}$			&&&&&	$0.9_{-0.5}^{+1.6}$				&&&								\\[0.5ex]	
		OBS3		&&&&		$4.6_{-1.2}^{+1.4}$			&&&&&	$0.12_{-0.08}^{+0.28}$			&&&								\\[0.5ex]	
		OBS4		&&&&			2.4 $^{\dag}$			&&&&&	$\leq1.2$						&&&								\\[0.5ex]	
	\end{tabular}	\\[0.8ex]
	\begin{tabular}{cccccccccccc}
		\midrule
										\multicolumn{12}{c}{\textbf{Model E:} Model D with the same intrinsic luminosity during the monitoring}			\\
		\midrule
					&&&&		$\rm{N_{H}^{l.o.s.}}$ [$10^{24}$ cm$^{-2}$]		&&&&		$\rm{norm_{intr}}$ [ph cm$^{-2}$ s$^{-1}$ keV$^{-1}$]			&&&		$\rm{\chi^2/dof}$		\\[0.5ex]
		OBS1		&&&& 	$7.4_{-1.0}^{+1.1}$			&&&&	\multirow{4}{*}{$0.5_{-0.2}^{+0.5}$}	&&&		\multirow{4}{*}{793/733}	\\[0.5ex]	
		OBS2		&&&& 	$5.6_{-0.8}^{+0.9}$			&&&&								&&&							\\[0.5ex]	
		OBS3		&&&&	$6.2_{-0.8}^{+1.0}$			&&&&								&&&							\\[0.5ex]	
		OBS4		&&&&		$\geq7.8$				&&&&								&&&							\\[0.5ex]	
	\end{tabular}	\\[0.8ex]
	\begin{tabular}{ccccccccccccc}
		\midrule
										\multicolumn{13}{c}{\textbf{Model F:} Model D with the same absorbing column density during the monitoring}		\\
		\midrule
		&			&&&&			$\rm{N_{H}^{l.o.s.}}$ [$10^{24}$ cm$^{-2}$]				&&&&		$\rm{norm_{intr}}$ [ph cm$^{-2}$ s$^{-1}$ keV$^{-1}$]			&&&		$\rm{\chi^2/dof}$		\\[0.5ex]
		&OBS1		&&&& 		\multirow{4}{*}{$5.9_{-0.8}^{+1.0}$}			&&&&		$0.15_{-0.10}^{+0.25}$		&&&		\multirow{4}{*}{793/733}	\\[0.5ex]	
		&OBS2		&&&& 											&&&&		$0.7_{-0.4}^{+0.8}$			&&&							\\[0.5ex]	
		&OBS3		&&&&											&&&&		$0.38_{-0.20}^{+0.47}$		&&&							\\[0.5ex]	
		&OBS4		&&&&											&&&&			$\leq0.14$				&&&							\\[0.5ex]	
	\bottomrule
	\end{tabular}	\\[0.8ex]
	\raggedright \textit{Notes.} \textit{Model A}. $\rm{\Gamma_{ULX}}$, $\rm{E_c}$, $\rm{norm_{ULX}}$: photon index, cut-off energy and normalization of the cut-off power-law accounting for the contribution of the ULX discussed in Appendix \ref{ULX}. \textit{Model B}. $\rm{E_{\texttt{gauss}}}$ and $\rm{norm_{\texttt{gauss}}}$: energy and normalization of the phenomenological line due to calibration issues modeled in XSPEC with $\sigma=0$ and discussed in Appendix \ref{cal_line}. \textit{Model C}. $\rm{\Gamma}$: photon index of the primary continuum; $\rm{N_{H,1}}$, $\rm{N_{H,2}}$, $\rm{N_{H,3}}$ and $\rm{norm_1}$, $\rm{norm_2}$, $\rm{norm_3}$: column densities and normalizations of the three reflectors, respectively.  \textit{Model D,E,F}. $\rm{N_{H}^{l.o.s.}}$: absorbing column density along the line of sight; $\rm{norm_{intr}}$: normalization of the intrinsic power law. $\rm{\chi^2/dof}$: ratio between $\chi^2$ and the degrees of freedom of the model. $^{\dag}$ Unconstrained value.
	   Errors correspond to the 90 per cent confidence level for one interesting parameter. In all models, ULX parameters are fixed to their best-fit values, while the column density along the line of sight and the intrinsic luminosity of the source are left free to vary (for clarity purposes, these values are not reported here for Model A,B,C). All the other parameters not shown in this table are fixed to the best-fit values reported in \citet{Bauer2015}.
\end{table*}

Moreover, to better compare our findings to those obtained from the previous \textit{NuSTAR} analyses, we decided to leave the column densities and normalizations of the reflection parameters free to vary (Model C - $\chi_r^2=1.08$ -- panel (c) in Figure \ref{fig:fit}).
This allowed us to check the good agreement between our values and those observed by \cite{Bauer2015} for the 2012 epoch and \cite{Marinucci2016} for the 2014-2015 epochs, as reported in Table \ref{tab:refl_free}.
Since the highest column density reflector located within 2 arcsec of the nuclear region and providing the bulk of the flux of the Compton hump (i.e. $N_{H,1}$) was not constrained, we fixed its value to the best-fit one (i.e. $10^{25}$ cm$^{-2}$).

\begin{table*}
	\centering
	\small
	\caption{Reflectors parameters modeled through MYTORUS tables \citep{MurphyYaqoob2009}.}
	\label{tab:refl_free}
	\begin{tabular}{cccccccc}
		\toprule
					&$\rm{\Gamma}$ $^{(1)}$	&$\rm{N_{H,1}}$ $^{(2)}$	&$\rm{norm_1}$ $^{(3)}$			&$\rm{N_{H,2}}$ $^{(2)}$	&$\rm{norm_2}$ $^{(3)}$			&$\rm{N_{H,3}}$ $^{(2)}$	&$\rm{norm_3}$ $^{(3)}$		\\
		\midrule
		Bauer (2015)	&$2.10_{-0.07}^{+0.06}$	&$\ge9.7$				&$(3.0\pm0.5)\times10^{-1}$		&$0.14\pm0.01$		&$(3.6_{-0.2}^{+0.3})\times10^{-2}$	&$5.0_{-1.9}^{+4.2}$		&$(1.0\pm0.2)\times10^{-2}$	\\[1.5ex]
		Marinucci (2016) 	&\multicolumn{7}{c}{all the parameters are fixed to the best-fit values found in Bauer (2015)} 																						\\[1.5ex]
This work (Model C)	 	&$2.10\pm0.01$	&$10^{\dag}$			&$(3.0\pm0.2)\times10^{-1}$			&$0.13\pm0.01$		&$(4.3_{-0.4}^{+0.3})\times10^{-2}$	&$\ge4.3$				&$(0.9\pm0.3)\times10^{-2}$	\\
		\bottomrule
	\end{tabular} \\
	\raggedright \textit{Notes.} $^{(1)}$ Photon index. $^{(2)}$ Column density of the reflector, in units of $10^{24}$ cm$^{-2}$.  $^{(3)}$ Normalization of the reflected emission, in units of ph cm$^{-2}$ s$^{-1}$ keV$^{-1}$. $^{\dag}$Fixed value.
\end{table*}

\begin{figure*}
	\centering
	\includegraphics[width=2.08\columnwidth]{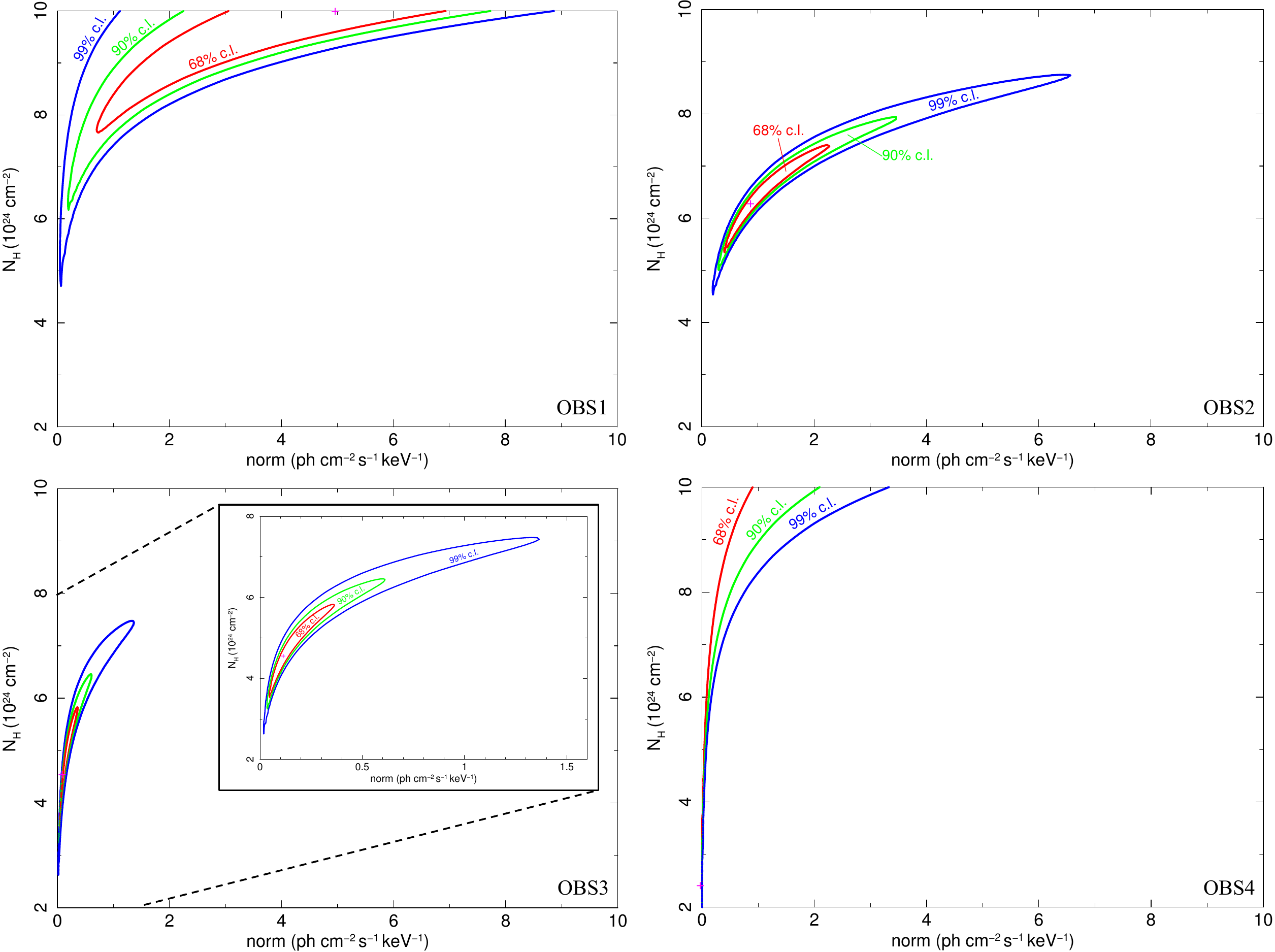}
	\caption{Confidence contours between the obscuring column density along the line of sight and the normalization of the primary power-law component for OBS1 (top left), OBS2 (top right), OBS3 (bottom left -- a zoom is shown within the black box), and OBS4 (bottom right), when Model D is assumed. A variation in $N_H$ is clearly visible between the observations, even if the absorbing column density is unconstrained in OBS4, while the normalizations are consistent with each other despite a large spread in their values. Red, green and blue contours indicate 68 per cent, 90 per cent and 99 per cent confidence levels, while the magenta cross indicates the best-fit values.}
    \label{fig:contour}
\end{figure*}

Then, due to the complexity of the model used to reproduce our data set (see Appendix \ref{model} for a detailed description), we decided to follow the same approach used by \cite{Marinucci2016} in analyzing the 2014-2015 \textit{NuSTAR} campaign.
Therefore, we fixed the column densities and the normalizations of the three reflectors to the best-fit values previously found, leaving only the flux of the primary component and the column density along the line of sight free to vary (Model D).
We obtained an improvement of the fit ($\chi_r^2=1.07$ -- panel (d) in Figure \ref{fig:fit}), pointing out a variation in $N_H$ between the observations.
However, the column density along the line of sight in OBS4 is unconstrained and a significant spread in the intrinsic emission of the AGN is clearly visible during the monitoring (see Figure \ref{fig:contour} and Model D in Table \ref{tab:models}), although the large errors make the normalizations of the four observations consistent with each other.

\begin{figure}
	\centering
	\includegraphics[width=1.0\columnwidth]{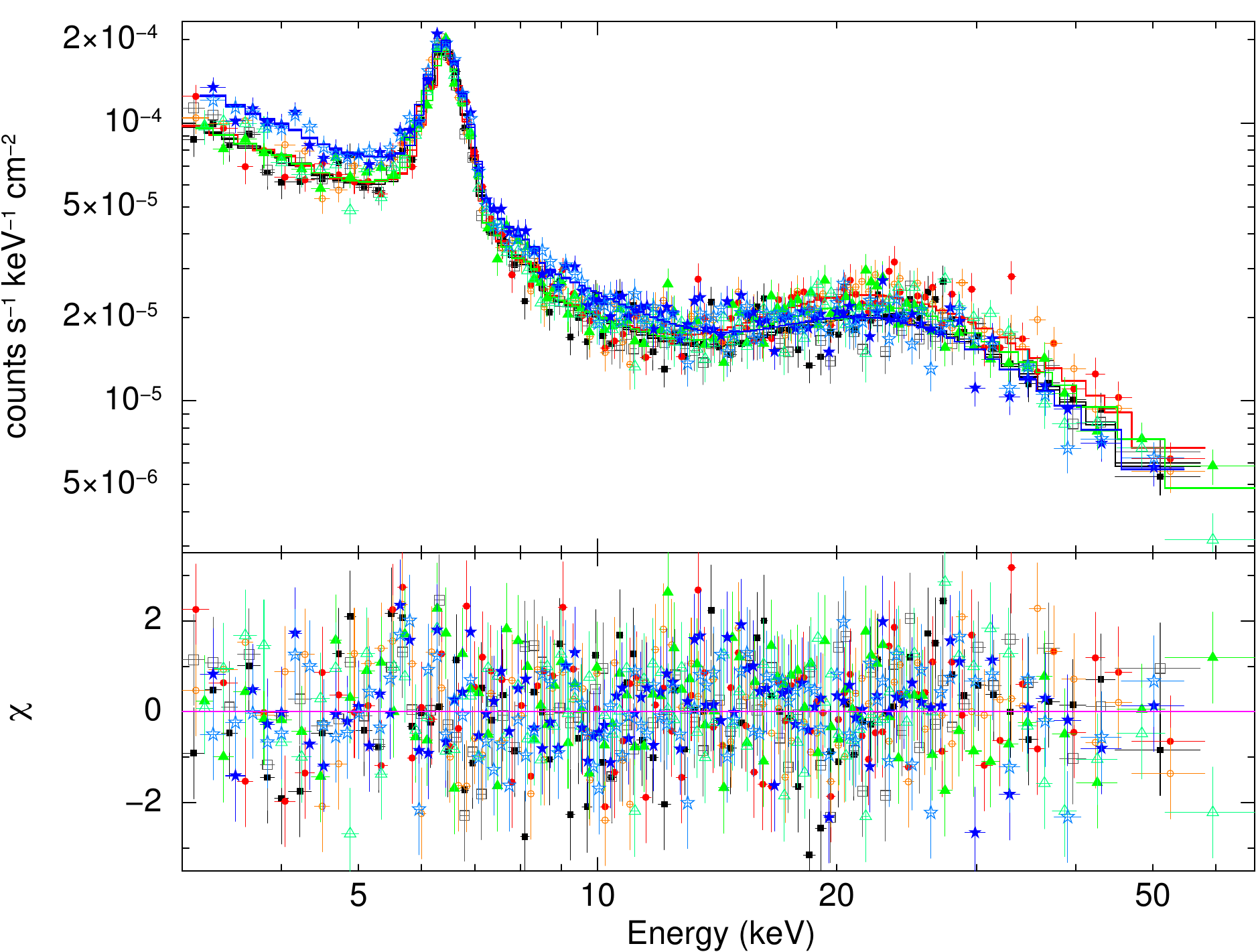}
	\caption{\textit{NuSTAR} monitoring spectra (top panel) and residuals plotted in terms of sigmas (bottom panel) with respect to the best-fit model (Model E). Both the color code and the symbols are the same as in Figure~\ref{fig:fit}.}
    \label{fig:bfit}
\end{figure}

For this reason, we tied together the normalizations of the intrinsic power-law during the whole monitoring, leaving only the column density along the line of sight free to vary (Model E - see Figure \ref{fig:bfit}).
Despite the $\chi_r^2$ being unchanged with respect to Model D, this modelling allowed us to avoid the degeneracy between variations in the absorbing column density and the intrinsic emission of the source resulting in better constraints on the column density in all the observations (see Model E in Table \ref{tab:models}) and reproducing well the behavior already shown in Figure \ref{fig:spectraHS-LS}.
Using this parametrization, we obtained an intrinsic unabsorbed 2-10 keV luminosity $L_{2-10}=(3.5_{-1.8}^{+3.6})\times10^{43}$ erg s$^{-1}$ at the distance of NGC 1068, in agreement with \cite{Bauer2015} and \cite{Marinucci2016}, and fully consistent with the one inferred using the mid-IR \citep{Gandhi2009} and [OIII] \citep{Lamastra2009} observed luminosities.

On the other hand, the degeneracy between $N_H$ and the intrinsic emission of the source could be also avoided assuming a uniform absorbing column density with an intrinsic variability in the accretion rate.
Testing this scenario (Model F in Table~\ref{tab:models}), we obtained a fit equivalent to the previous one from a statistical point of view ($\chi^2_r=1.08$), with an absorbing column density along the line of sight $N_H=(5.9_{-0.8}^{+1.0})\times10^{24}$ cm$^{-2}$, fully consistent with the values obtained in Model D only for two observations (OBS2 and OBS3).
Furthermore, a variation between the normalizations of the intrinsic power-law in OBS3 and OBS4 is clearly visible.
This corresponds to a change in the X-ray luminosity in the 2-10 keV band $\Delta L_{2-10}=(1.8_{-1.4}^{+3.3}\times10^{43})$ erg s$^{-1}$, suggesting a decrease in the normalized accretion rate of the source (i.e. $\rm L_{bol}/L_{Edd}$) from 0.47 to 0.12 in a 3-month timescale.

However, due to the fact that the X-ray luminosity we obtain in the low-flux state of NGC 1068 (which characterises most of the observations) using Model F (i.e. $L_{2-10}\leq~9.8\times10^{42}$ erg s$^{-1}$) is much less than the one expected from other wavelengths, while the constant $L_X$ - multiple $N_H$ scenario provides an X-ray luminosity fully consistent with the one inferred using the mid-IR and [OIII] observed luminosities, we adopted Model E as our best-fit model.

\section{Discussion}
\label{Section4}

\begin{figure}
	\centering
	\includegraphics[width=1.0\columnwidth]{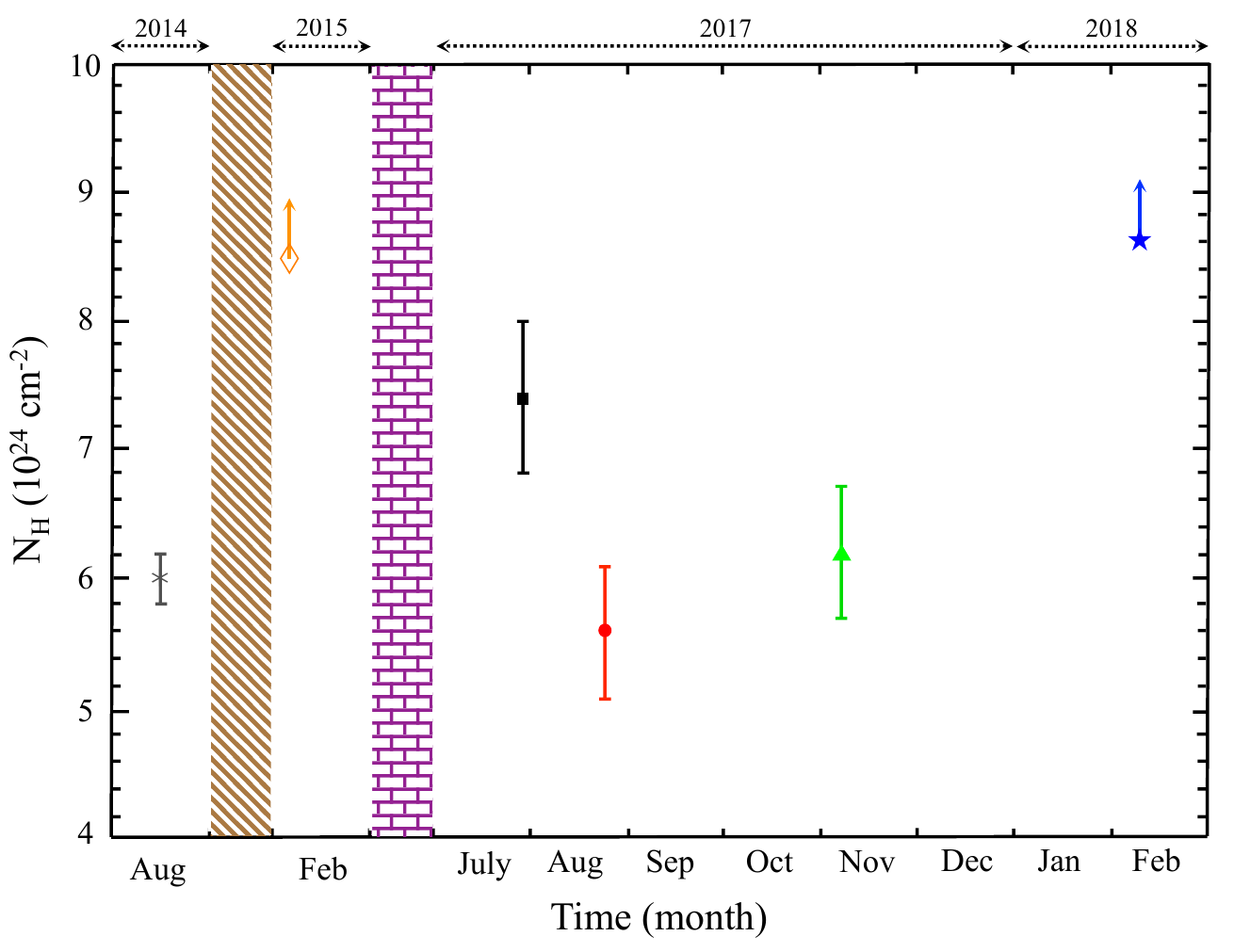}
	\caption{Values of the obscuring column density along the line of sight for the four 2017-2018 NuSTAR monitoring observations (black square, red circle, green triangle and blue star), when Model E is adopted. Old 2014-2015 measurements (grey asterisk and orange diamond) reproducing the same intrinsic luminosity are extrapolated from Figure 3 in \citet{Marinucci2016} and shown here for comparison. Errors correspond to the 68 per cent confidence level for one interesting parameter. Brown stripes and violet bricks indicate a period of 5 and 28 months, respectively, not covered by observations.}
    \label{fig:plot}
\end{figure}

Supposing that the intrinsic luminosity of the source did not vary during the whole monitoring, we attribute the spectral differences observed above 20 keV to X-ray emission piercing through a patchy dusty region.
We observe the same column density variability on 6-month timescales already found by \citet{Marinucci2016} during the previous \textit{NuSTAR} monitoring campaign (see red and blue data points with respect to grey and orange ones in Figure \ref{fig:plot}), suggesting the presence of a variable absorber on parsec-scale distance.
However, the observational strategy of our monitoring campaign allowed us to probe shorter timescales with the aim of better constraining the location of the circumnuclear absorbing matter in NGC 1068.

	According to our best-fit model, the values of the column density along the line of sight with their 1-sigma errors are $N_H=(7.4\pm0.6)\times10^{24}$ cm$^{-2}$, $N_H=(5.6\pm0.5)\times10^{24}$ cm$^{-2}$, $N_H=(6.2\pm0.5)\times10^{24}$ cm$^{-2}$ and $N_H\geq8.6\times10^{24}$ cm$^{-2}$ (corresponding to $N_H\geq7.8\times10^{24}$ cm$^{-2}$ at the 90 per cent confidence level, as reported in Table \ref{tab:models}) for OBS1, OBS2, OBS3 and OBS4, respectively (see Figure \ref{fig:plot}).
	Thus, assuming the same intrinsic X-ray luminosity during the whole monitoring, we could clearly identify one unveiling and one eclipsing event: the first between July and August 2017 and the second between November 2017 and February 2018 with timescales lower than 27 days and 91 days, respectively.
	These events were ascribed to Compton-thick material with $N_H=(1.8\pm0.8)\times10^{24}$ cm$^{-2}$ and $N_H\geq(2.4\pm0.5)\times10^{24}$ cm$^{-2}$, respectively, which moved temporarily across our line of sight, allowing or preventing the intrinsic nuclear radiation to pierce through the circumnuclear absorbing medium.	

	Considering a scenario in which the obscuring material is composed of individual spherical clouds orbiting with Keplerian velocities at a distance $R$ from the SMBH, the distance of the cloud responsible for the unveiling/eclipsing event is given by
\begin{equation}
	\rm R[cm]=\frac{GM_{BH}(\Delta t)^2n^2}{N_H^2},
	\label{eq:R}
\end{equation}
where $G$ is the gravitational constant, $M_{BH}$ is the BH mass, $\Delta t$ is the timescale of the column density variation, and $n$ and $N_H$ are the gas density and the column density of the cloud, respectively.
Assuming a black hole mass $M_{BH}=10^7$ M$_{\odot}$, as derived from water maser measurements \citep{Greenhill1996}, and a typical value for the gas density within the broad line region, such as $n=10^{10}$ cm$^{-3}$ \citep{Wang2012}, we estimate a size for the obscuring cloud of $D\sim150\ R_g$ and find that the absorbers responsible for the unveiling and eclipsing event are located at a distance $R\leq(0.07\pm0.05)$ pc and $R\leq(0.46\pm0.14)$ pc, respectively.

According to \citet{Kaspi2005}, the radius of the BLR is given by
\begin{equation}
	\rm\frac{R_{BLR}}{10\ lt-days}=0.86\times\left(\frac{L_{2-10\ keV}}{10^{43}\ erg\ s^{-1}}\right)^{0.544},
	\label{eq:BLR}
\end{equation}
leading to $R_{BLR}=\left(0.014_{-0.007}^{+0.015}\right)$ pc, using our spectral fitting parameters.
Moreover, using the bolometric correction from \cite{Marconi2004}, we inferred a bolometric luminosity $L_{bol}=(0.8_{-0.6}^{+1.2})\times10^{45}$ erg s$^{-1}$, in agreement with other studies (e.g. \citealp{Pier1994}; \citealp{WooUrry2002}; \citealp{Honig2008}) and leading to a dust sublimation radius $R_d=(0.36_{-0.14}^{+0.27})$ pc, if an average dust grain size of 0.05 $\mu$m with a temperature $T=1500\ K$ is considered \citep{Barvainis1987}.
Assuming the torus inner walls to be at the dust sublimation radius, this value is consistent with the dimension of the dusty torus observed in infrared.
Therefore, considering the gas density reported above, the obscuring clouds are constrained to be located in the innermost and hottest part of the dusty torus, or even further inside (i.e. in the BLR).

Finally, the inferred bolometric luminosity leads to a normalized accretion rate $\lambda\color{black}=\frac{L_{\rm bol}}{L_{\rm Edd}}=0.63_{-0.48}^{+0.95}$\color{black}, which confirms the highly accreting nature of the source.
Therefore, we cannot definitively rule out the hypothesis that at least part of the observed X-ray variability of the AGN in NGC 1068 is due to a change in the intrinsic luminosity of the central accretion disk.
But, also in this latter case (i.e. Model D), a variation in $N_H$ is expected, even though the column density in OBS4 is unconstrained.
On the other hand, if we attributed the observed X-ray variability as due only to a change in the intrinsic AGN luminosity (i.e. Model F), we obtain a scenario equivalent to that suggested by our best-fit model from a statistical point of view.
However, according to \citet{Shemmer2008}, the normalized accretion rate resulting in OBS4 would correspond to a power-law slope $\Gamma=1.82\pm0.01$, well below that observed in our analysis (i.e. $\Gamma=2.1$).
Furthermore, the uniform $N_H$ - multiple $L_X$ scenario predicts a column density well below the average value of 10$^{25}$ cm$^{-2}$ usually observed in NGC 1068 (e.g. \citealp{Bauer2015}), while our assumption of a constant intrinsic luminosity of the AGN on timescales of months is supported by mid-IR and optical data.
Thus, our findings firmly rule out the scenario of a single, monolithic obscuring wall, and instead support the presence of a clumpy torus surrounding the nuclear region of NGC 1068.

\section{Conclusions}
\label{Section5}
We have presented a spectral analysis of the latest \textit{NuSTAR} monitoring campaign of the Compton-thick Seyfert 2 galaxy NGC 1068 composed of four $\sim$50 ks observations performed between July 2017 and February 2018 and probing timescales from 1 to 6 months.
Our findings are summarized as follows.

\begin{itemize}
	\item We detected one unveiling and one eclipsing event with timescales lower than 27 and 91 days, respectively, ascribed to Compton-thick material with $N_H=(1.8\pm0.8)\times10^{24}$ cm$^{-2}$ and $N_H\geq(2.4\pm0.5)\times10^{24}$ cm$^{-2}$ moving across our line of sight, allowing or preventing the intrinsic nuclear radiation from piercing through the circumnuclear matter.
	We can apparently locate the absorbing gas clouds to arise from the innermost part of the torus or closer, thus providing both tighter constraints on their location with respect to the previous \textit{NuSTAR} monitoring campaign, and further evidence of the clumpy structure of the circumnuclear matter in this source.
	\item We reported a \textit{Swift}-XRT detection of a transient X-ray source in February 2018, which is plausibly a strongly variable ULX located at $\sim$2 kpc from the nucleus of NGC 1068, with a peak X-ray intrinsic luminosity of $(3.0\pm0.4)\times10^{40}$ erg s$^{-1}$ in the 2-10 keV band.
\end{itemize}

	Since its launch in June 2012, \textit{NuSTAR} has observed NGC 1068 several times, sampling timescales ranging from $\sim$1 month to $\sim$5 years; further observations also performed simultaneously with future X-ray observatories, like \textit{XRISM}, will be certainly helpful in refining and better constraining the scenario we discussed in this paper.
	Furthermore, a leap in this field is expected to be achieved with the launch of the next future High-Energy X-ray Probe (HEX-P -- \citealp{Madsen2019}) taking advantage of the complementary capability with the simultaneous  \textit{ATHENA} mission (\citealp{Nandra2013}; \citealp{Barcons2017}), providing the former a high-energy sensitivity with an improvement by a factor of 40 over \textit{NuSTAR} in the 10-80 keV band, and the latter a high resolution spectroscopy below 10 keV.
On the other hand, different information on the nature and geometry of the circumnuclear matter in NGC 1068 can be obtained from future X-ray polarimetry missions, such as the Imaging X-ray Polarimeter Explorer (\textit{IXPE} - \citealp{Weisskopf2016a} and \citealp{Weisskopf2016b}, scheduled to be launched in 2021) and the enhanced X-ray Timing and Polarimetry mission (\textit{eXTP} -- \citealp{Zhang2019}, planned to be launched in 2027).
In fact, in the 2-8 keV working band of the polarimeters on board these missions, the X-ray emission is dominated by reflection from the circumnuclear matter.
A high polarization degree, dependent on the inclination and level of symmetry of the matter, is expected, with the polarization angle related to the symmetry axis.
A comparison of such an axis to those of the spatially resolved inner tori and of the ionization cone will shed further light on the structure of the circumunuclear matter in NGC 1068.

\section*{Acknowledgements}
We thank the anonymous referee for her/his useful comments and suggestions which improved the quality and the clarity of the paper.
AZ, SB and GM acknowledge financial support from the Italian Space Agency under grant n. 2017-14-H.O.
SB acknowledges financial support from the Italian Space Agency under grant ASI-INAF I/037/12/0.
AM and GM acknowledge financial support from the Italian Space Agency under grant ASI-INAF I/037/12/0-011/13.
FEB acknowledges support from CONICYT-Chile (Basal AFB-170002) and the Ministry of Economy, Development, and Tourism's Millennium Science Initiative through grant IC120009, awarded to the Millennium Institute of Astrophysics, MAS.
CR acknowledges support from the CONICYT+PAI Convocatoria Nacional subvencion a instalacion en la academia convocatoria a\~{n}o 2017 PAI77170080.
This research made use of data from the \textit{NuSTAR} mission, a project led by the California Institute of Technology, managed by the Jet Propulsion Laboratory, and funded by NASA, and of the \textit{NuSTAR} Data Analysis Software (NuSTARDAS) jointly developed by the ASI Science Data Center (ASDC, Italy) and the California Institute of Technology (Caltech, USA).
We also thank the \textit{Swift} team for having scheduled our ToO request as soon as possible, and provided useful data for this paper through archival XRT observations.


\appendix

\section{A new ULX detection in NGC 1068?}
\label{ULX}

NGC 1068 is not only one of the best known Seyfert 2 galaxies in the local universe, but also a powerful starburst galaxy with a star-formation rate $\dot M>36\ M_{\odot}\ yr^{-1}$ (i.e. at least a factor of 7 greater than the SFR in our Galaxy).
In particular, infrared observations indicate that the star formation in this galaxy is primarily distributed within the central $\sim$3 kpc in two very extended complexes, one located to the north and the other to the southwest of the nucleus \citep{TelescoDecher1988}.
	
\begin{figure}
	\centering
	\includegraphics[width=1.0\columnwidth]{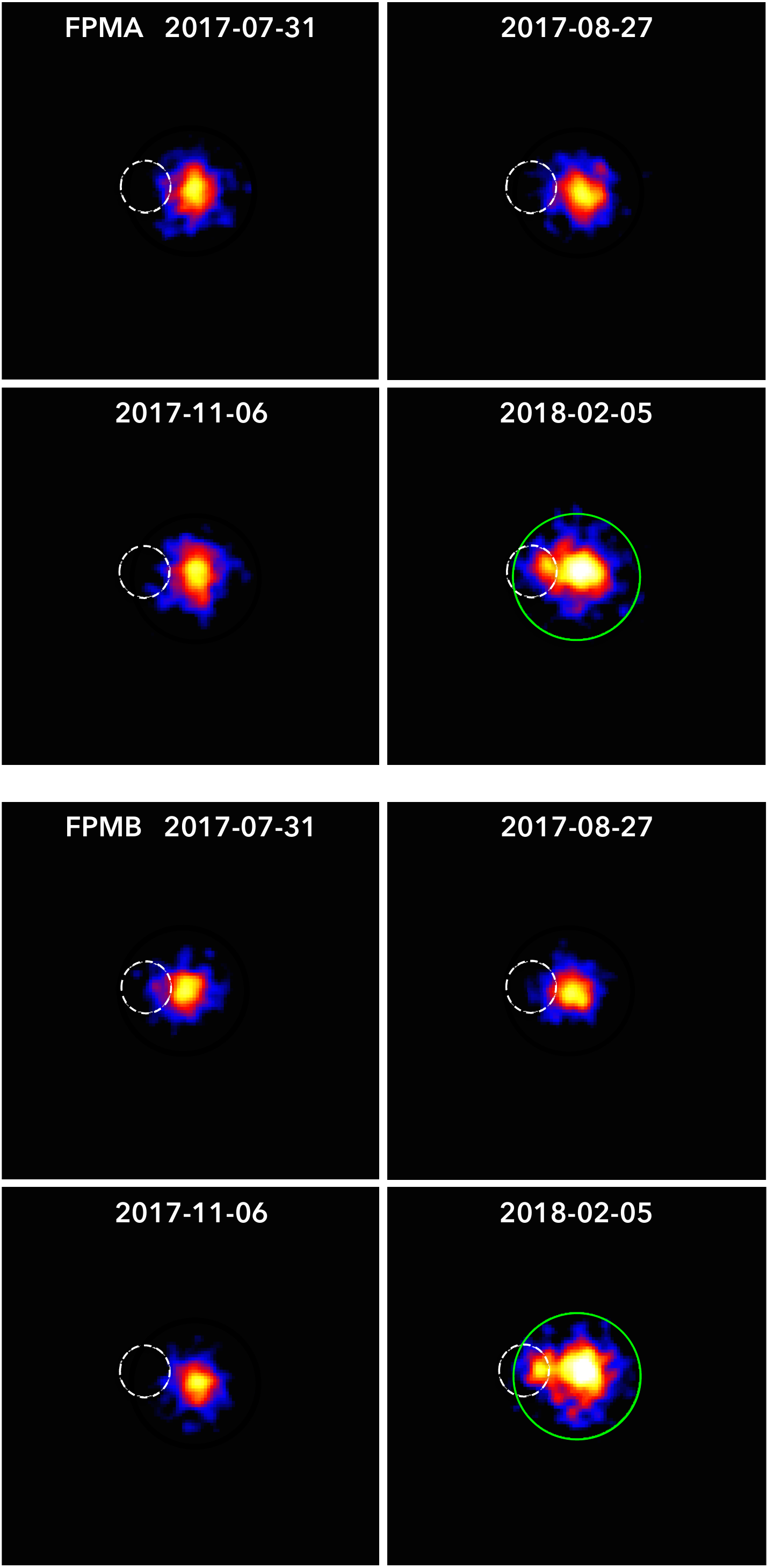}
	\caption{\textit{NuSTAR} FPMA (top panels) and FPMB (bottom panels) images of the central 5'$\times$5' regions in the 3-5.5 keV band smoothed with a 3 pixels radius Gaussian filter. The circular region with a radius of 50 arcsec used for extracting the source spectra is shown in OBS4 (green solid circle), while the 20 arcsec circular region used to extracted the \textit{Swift}-XRT spectrum is overplotted in all panels (white dotted circles).}
	\label{fig:ULX_nu}
\end{figure}

\begin{figure}
	\centering
	\includegraphics[width=0.967\columnwidth]{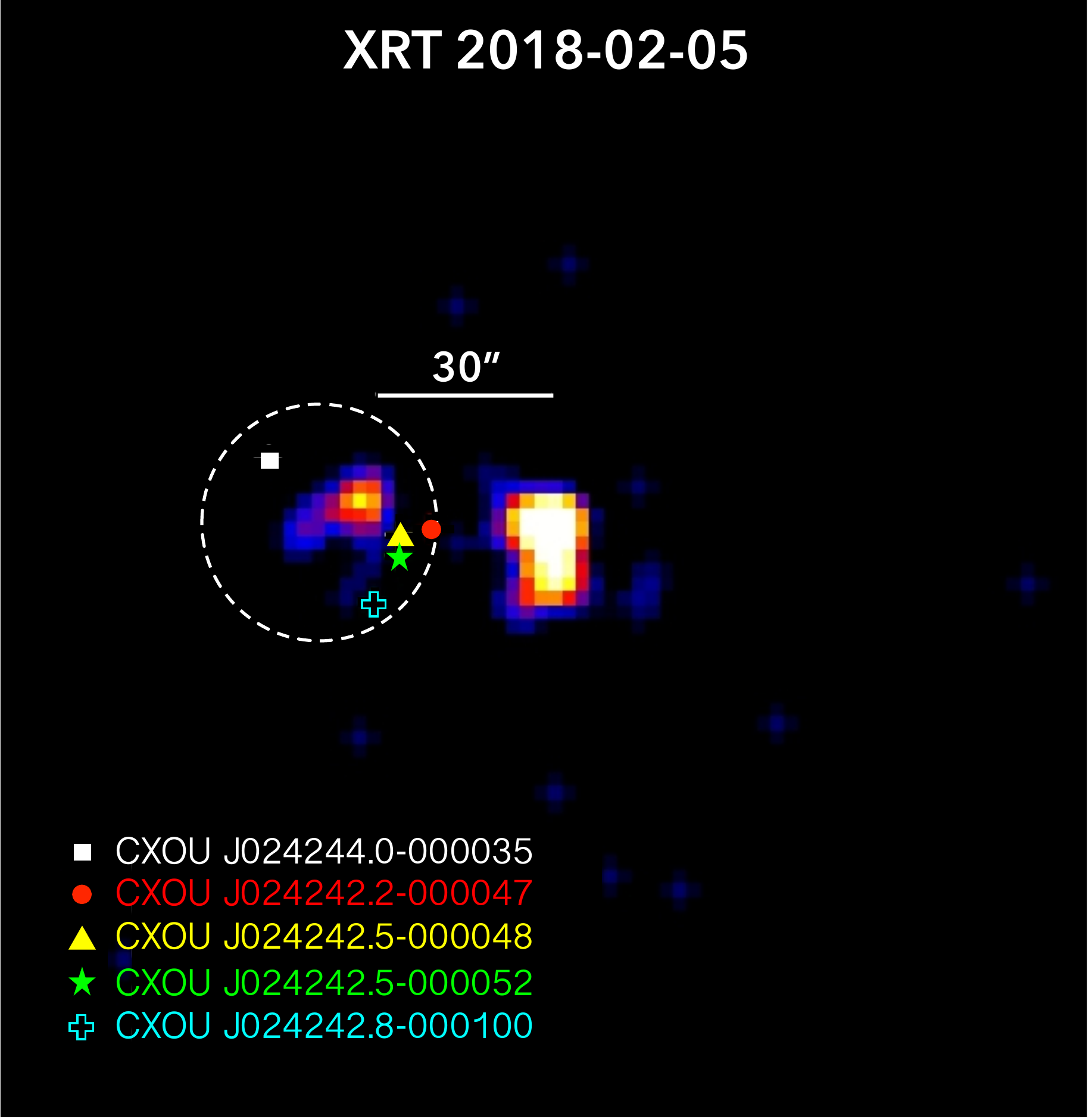}
	\caption{\textit{Swift}-XRT image of the central 3'$\times$3' region, obtained in February 2018, in the 3-5.5 keV band. The dashed white circular region with a radius of 20 arcsec used for extracting the ULX spectrum is shown. We over-imposed the five different point-like sources previously detected with \textit{Chandra} and located within a 20 arcsec radius region from the ULX, from \citet{SmithWilson2003}. The ULX is not detected in November 2017 nor in June 2018.}
	\label{fig:ULX_sw}
\end{figure}

	The \textit{NuSTAR} images taken in February 2018 in the 3-5.5 keV energy band suggest the presence of a transient object at a distance of $\sim$30 arcsec from the nuclear region (see Figure \ref{fig:ULX_nu}).
Analysis of the simultaneous \textit{Swift}-XRT image confirms this hypothesis (see Figure \ref{fig:ULX_sw}), allowing us to ascribe the 3-5.5 keV excess observed in February 2018 to the appearance of a flaring source at a distance of $\sim$2 kpc from the nucleus of NGC 1068 (at the distance of NGC 1068 1"=72 pc -- \citealp{Bland-Hawthorn1997}).

To identify this flaring source, we overlaid the five point-like sources previously detected with \textit{Chandra} \citep{SmithWilson2003}, within 20 arcsec of its position, on the \textit{Swift}-XRT 3-5.5 keV image (see Figure \ref{fig:ULX_sw}).
Among these point-like objects, the brightest one (with more than 50 counts in the \textit{Chandra} band and with a SNR>7) is CXOU J024244.0-000035, with a reported 0.4-8 keV luminosity $L_{0.4-8}=6.8\times10^{38}$ erg s$^{-1}$, corresponding to a 2-10 keV luminosity $L_{2-10}=4.8\times10^{38}$ erg s$^{-1}$, which is almost two orders of magnitude lower than the one observed during our monitoring.
No known X-ray sources appear to lie closer than $d\simeq8$ arcsec from the ULX centroid in the \textit{Swift}-XRT image, neither the supernova exploded in NGC 1068 in November 2018 \citep{Bostroem2019} matches the location of the ULX; therefore, we conclude that there is no robust identification with previously detected sources.

To obtain \textit{NuSTAR} FPMA and FPMB spectra of the ULX, we computed difference spectra between the ones extracted from the February 2018 and November 2017 observations, already described in Section \ref{NuSTAR_obs}.
Then, we rebinned the two spectra in order to have a SNR greater than 2$\sigma$ in each spectral channel. With this requirement, no detected spectral bins are present above 16 keV and 12 keV for FPMA and FPMB, respectively.

\begin{figure}
	\centering
	\includegraphics[width=1.0\columnwidth]{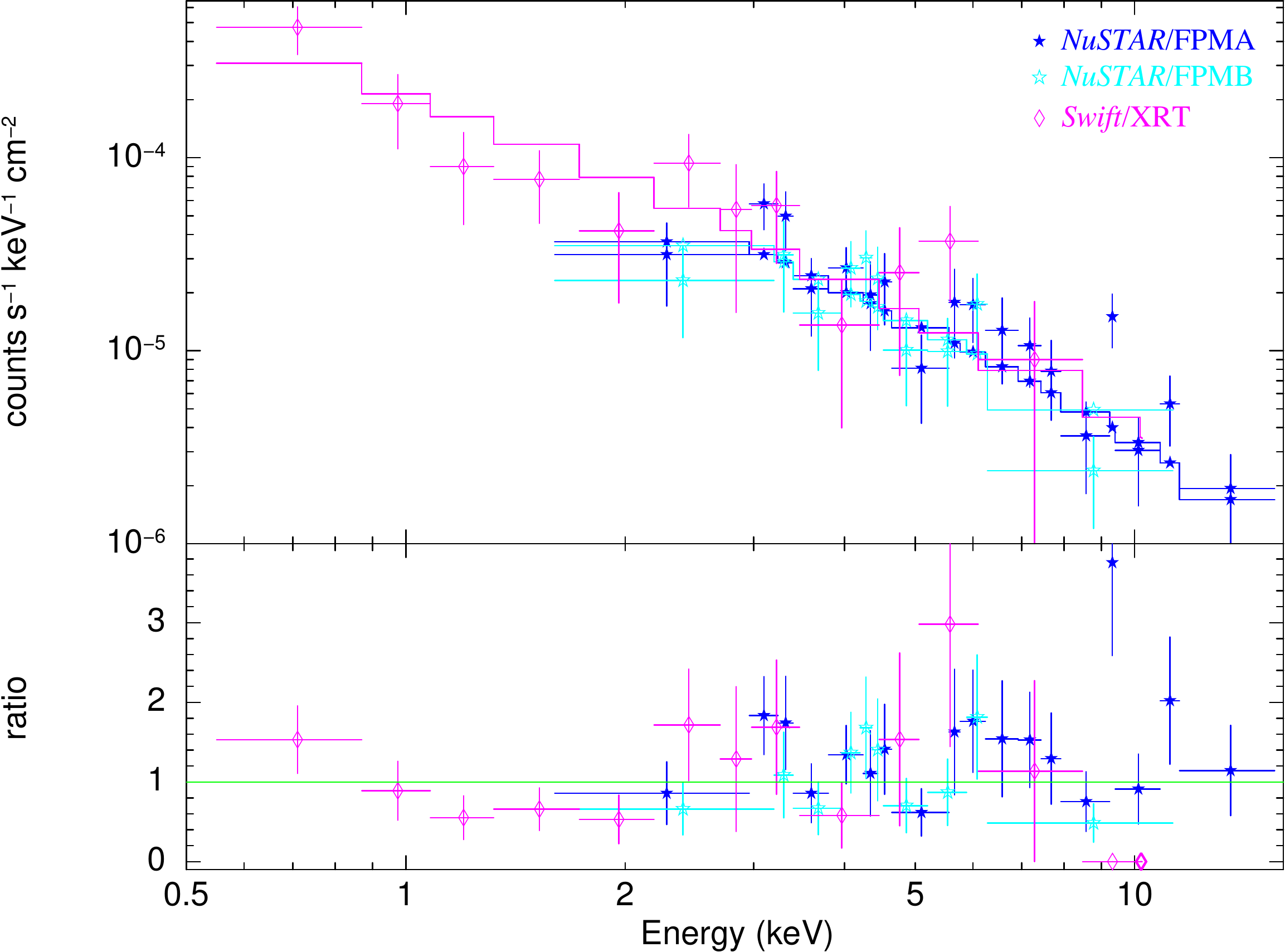}
    \caption{\textit{Top panel.} XRT (magenta diamonds), FPMA (blue filled stars) and FPMB (cyan empty stars) ULX spectra taken in February 2018. \textit{Bottom panel.} Data/model ratio with respect to a model (solid green line) composed of an absorbed power-law with a photon index $\Gamma=1.5\pm0.4$ and a cut-off energy $E_c\geq5$ keV.}
    \label{fig:Swift_spectra}
\end{figure}

We used the C-statistic to simultaneously fit the XRT, FPMA and FPMB spectra obtained in February 2018.
An absorbed cut-off power-law provides an acceptable description of the data, with a fixed value for the Galactic column density ($N_H^{Gal}=3\times 10^{20}$ cm$^{-2}$ -- \citealp{Kalberla2005}), a photon index $\Gamma=1.5\pm0.4$, a cut-off energy $E_c\geq5$ keV and a normalization $N=(2.6_{-0.6}^{+0.7})\times10^{-4}$ ph cm$^{-2}$ s$^{-1}$ keV$^{-1}$.
We retrieved a $C/dof=43.3/45$ and our best fit is shown in Figure \ref{fig:Swift_spectra}.
The observed flux in the 2-10 keV band is $F_{2-10}=(9.4\pm1.0)\times10^{-13}$ erg cm$^{-2}$ s$^{-1}$, corresponding to an intrinsic luminosity $L_{2-10}=(3.0\pm0.4)\times10^{40}$ erg s$^{-1}$ at the distance of NGC 1068.

Since the source was not detected before, we requested and obtained a 1.5 ks ToO observation with \textit{Swift}-XRT on June 15, 2018 (NGC 1068 was not observable before this date due to Sun occultation) to confirm the presence of this new source.
Unfortunately, the transient source was no longer visible in the \textit{Swift}-XRT image in the 3-5.5 keV band, and the spectrum extracted from a circular region of 20 arcsec centered on the position of the ULX, as observed in February 2018, showed a dramatic drop of the count-rate in the 2-10 keV band (i.e. about one order of magnitude, from $0.010\pm0.002$ counts/s in February 2018 to $0.0013\pm0.0009$ counts/s in June 2018).
This prevented further detailed study of this transient source, allowing us to determine only a variability timescale less than $\sim3$ months.
However, we note that transient ULXs are already known (e.g. \citealp{Middleton2012}, \citealp{Middleton2013}; \citealp{Pintore2018}), and that even high luminosity ULXs can vary on very short timescales, such as ULX1 in NGC 5907 \citep{Walton2015}.

\section{The Bauer model}
	\label{model}

In 2015, \citet{Bauer2015} characterized the multi-epoch X-ray spectra of NGC 1068 analyzing high-quality observations performed from 1996 until 2012 by different X-ray observatories.
In particular, \textit{NuSTAR} and \textit{XMM-Newton} data were used to study in detail the total emission of the source, while the higher spectral and angular resolution of \textit{Chandra} (HETG and ACIS, respectively) enabled the authors to remove potential host-galaxy contamination, spatially separating the nuclear spectrum from diffuse and off-nuclear point-source emission at least below 8 keV.
\textit{Swift}, \textit{Suzaku} and \textit{BeppoSAX} observations were used for points of comparison.

The broadband emission of NGC 1068 was modeled with a combination of a heavily Compton-thick transmitted power law, scattering by both warm and cold reflectors, radiative recombination continuum (RRC) and line emission, and off-nuclear point-source emission (see Figure 12 in \citealp{Bauer2015} for the fractional contributions of the different spectral components).
No angular dependence of the nuclear emission spectral shape is assumed (i.e. all scattering component have the same photon index), nor is relativistic reflection from the accretion disk considered, due to the inclination of the source and the dominance of scattering and absorption from cold distant material.

To construct a robust model for the nuclear X-ray spectrum of NGC 1068, both the point-like nuclear emission and the diffuse emission and point source contamination from the host galaxy must be taken into account.
The extended emission can be effectively constrained below 8 keV by \textit{Chandra} imaging, leaving considerable degeneracy at higher energies.
Since the aim is to provide the best constraints on the properties of the reflectors, data below 2 keV are not considered, and all the spectral components that are well-constrained through a separate fit of the nuclear and host galaxy contribution (i.e. extranuclear point sources, RRC and line emission) are fixed during the combined fit (see \citealp{Bauer2015} for the adopted values of each spectral component).

\begin{figure}
	\centering
	\includegraphics[width=1.0\columnwidth]{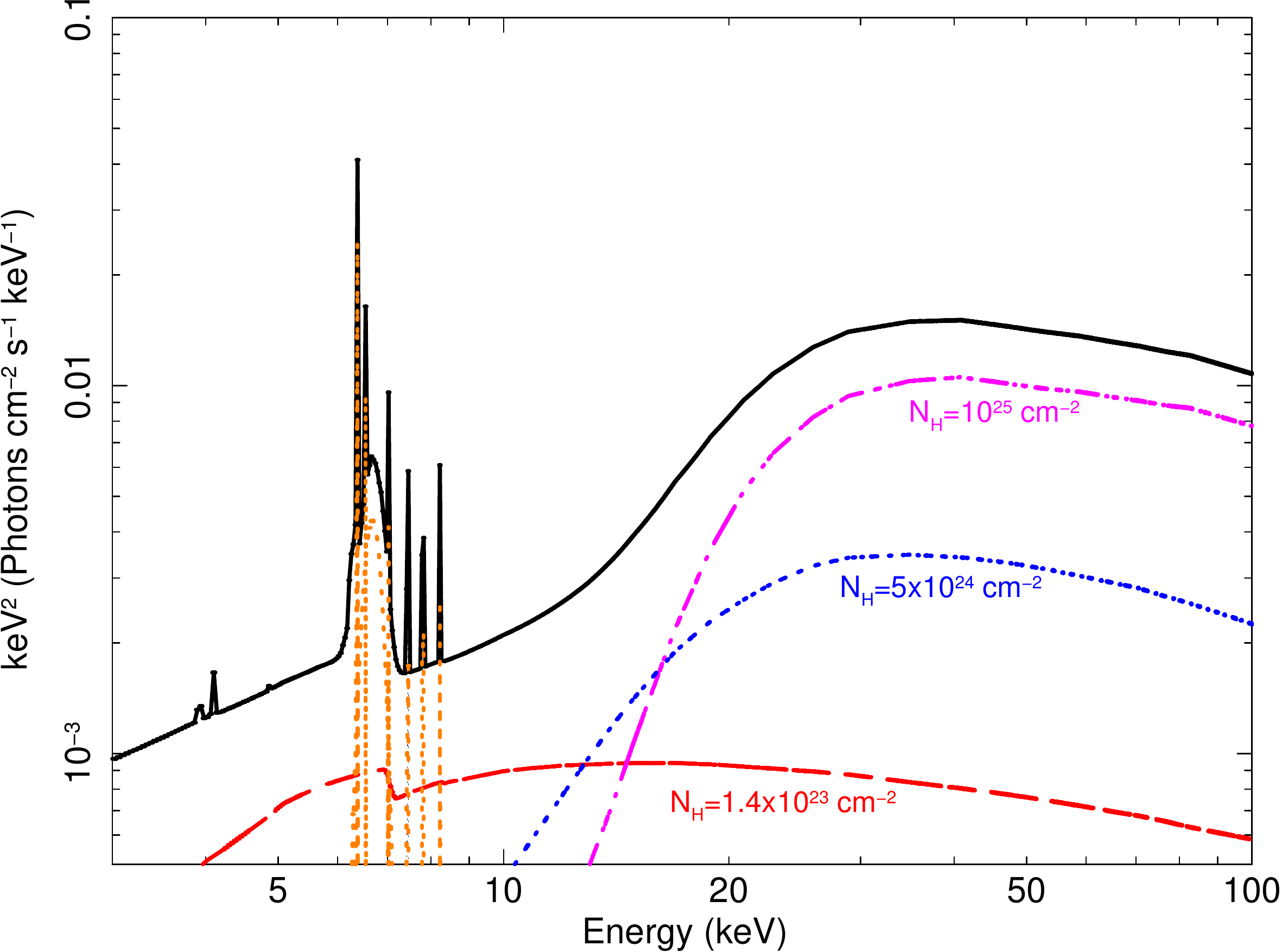}
   \caption{Theoretical best-fit model of NGC 1068 adopted by \citealp{Bauer2015} (solid black line). Emission lines (orange dotted lines in the 6-9 keV energy range), nuclear (dashed red and dot-dashed magenta lines) and more distant (blue dotted line) reflectors represent the reflection components arising from different column densities, while the nuclear intrinsic emission is completely absorbed by heavily Compton-thick matter.}
    \label{fig:Bauer_model}
\end{figure}

The AGN intrinsic continuum is well-described by a highly absorbed ($N_H=10^{25}$ cm$^{-2}$) power-law with a photon index $\Gamma=2.10_{-0.07}^{+0.06}$ and an energy cut-off $E_c=128_{-44}^{+115}$ keV, while a multi-component reflector with three distinct column densities ($N_{H,1}\sim10^{25}$ cm$^{-2}$, $N_{H,2}=(1.4\pm0.1)\times10^{23}$ cm$^{-2}$, $N_{H,3}=(5.0_{-1.9}^{+4.2})\times10^{24}$ cm$^{-2}$) reproduces the complex structure of the circumnuclear matter (i.e. Model M2d in \citealp{Bauer2015} -- see Figure \ref{fig:Bauer_model}).
The spectral features attributed to the $N_{H,1}$ and $N_{H,2}$ components arise from the central region, within 2 arcsec from the nucleus, while $N_{H,3}$ corresponds to regions outside the central 2 arcsec.
In particular, the higher $N_H$ component provides the bulk of the flux to the Compton hump, while the lower $N_H$ component contributes primarily to the iron line emission and reproduces the curvature of the continuum around 10 keV, effectively decoupling the two key features of Compton reflection.
The inclination angles of the two nuclear scatterers with respect to the line of sight are $\ang{90}$ and $\ang{0}$, respectively, in order to reproduce a clumpy torus distribution with the edge-on scatterer accounting for the photons reprocessed by the obscuring material lying between the AGN and the observer, while the face-on scatterer mimics the reprocessed emission coming from back-side reflection.
These latter photons have a smaller chance of being absorbed before reaching the observer, making the MYTORUS $\ang{0}$ component relevant in a patchy torus scenario (\citealp{Yaqoob2015}; \citealp{Zhao2019}).
On the other hand, the third reflector on more extended scales (>140 pc) with an inclination of $\ang{0}$ provides almost 30\% of the neutral iron K$\alpha$ line flux and could represent material within the ionization cones.

\section{Calibration issues}
\label{cal_line}

\begin{figure*}
	\centering
	\includegraphics[width=2.1\columnwidth]{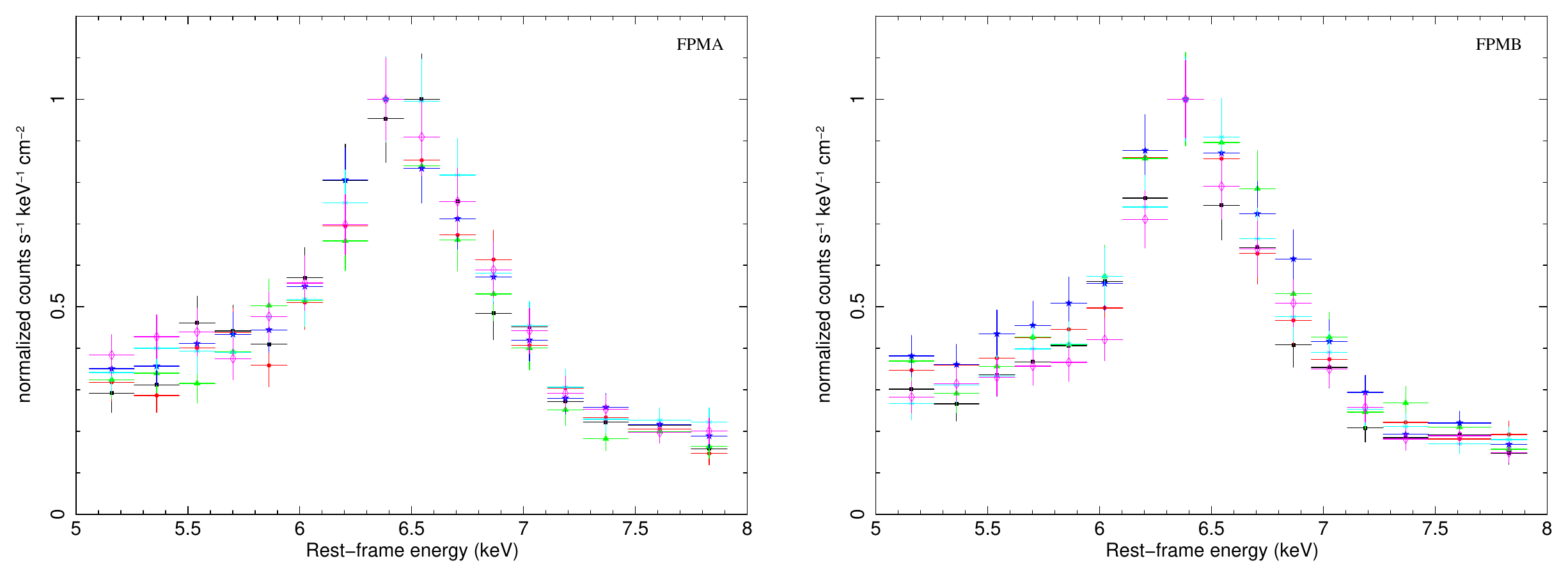}
	\caption{FPMA (left panel) and FPMB (right panel) spectra of NGC 1068 in the 5-8 keV range, normalizing to one the intensity of the highest channel. Black squares, red circles, green triangles, blue stars, cyan asterisks and magenta diamonds represent OBS1, OBS2, OBS3, OBS4, August 2014 and February 2015 observations, respectively. For clarity purposes, all the observations have the same energy bins.}
    \label{fig:NuSTAR_old/new}
\end{figure*}

As reported in Section \ref{Section3}, when modelling the \textit{NuSTAR} X-ray spectra of NGC 1068, we obtained significant residuals at $\sim$6 keV (see panel (a) in Figure \ref{fig:fit}).
Comparing our monitoring observations with older data obtained during the previous \textit{NuSTAR} monitoring campaign in 2014-2015, we find lack of temporal dependence of the iron line profile (see Figure \ref{fig:NuSTAR_old/new}).
Assuming that the slight variations of the blue data are due to the appearence of the ULX in February 2018 (see Appendix \ref{ULX}), the line profiles are sufficiently similar to each other to conclude that their variance is purely statistical.
We note that no issues were reported in \citet{Marinucci2016} because they only considered \textit{NuSTAR} data above 8 keV, using simultaneous \textit{XMM-Newton} data at lower energies.

Fitting the residuals at $\sim$6 keV with a further emission line (i.e. a \texttt{gauss} component in XSPEC) with respect to Model A, we obtain energies and normalization values fully consistent with each other, both in FPMA and FPMB (see Table \ref{tab:line}).
We note that for the data reduction of the older \textit{NuSTAR} observations we used the same procedure explained in Section \ref{NuSTAR_obs}, binning the data using OBS2 as a template (i.e. the August 2014 and February 2015 spectra have the same energy bins of the August 2017 one).

\begin{table}
	\centering
	\small
	\caption{Comparison of the 6 keV line component between old and new NuSTAR observations.}
	\label{tab:line}
	\begin{tabular}{lcc}
		\hline
						&	FPMA$^{\dag}$		&	FPMB$^{\dag}$		\\
		\hline
		Aug 2014			&	$1.8\pm0.5$		&	$1.8\pm0.5$		\\
		Feb 2015			&	$2.0\pm0.4$		&	$2.0\pm0.5$		\\
		OBS1 (Jul 2017)	&	$1.7\pm0.5$		&	$2.3\pm0.5$		\\
		OBS2 (Aug 2017)	&	$1.3\pm0.5$		&	$2.4\pm0.5$		\\
		OBS3 (Nov 2017)	&	$2.0\pm0.5$		&	$2.1\pm0.5$		\\
		OBS4 (Feb 2018)	&	$2.1\pm0.5$		&	$2.1\pm0.5$		\\
		\hline
	\end{tabular} \\
	\raggedright \textit{Note.} $^{\dag}$ Normalization of the spurious line component, in units of 10$^{-5}$ ph cm$^{-2}$ s$^{-1}$ keV$^{-1}$, fitting the data with Model E.
\end{table}

\begin{figure}
	\centering
	\includegraphics[width=1.0\columnwidth]{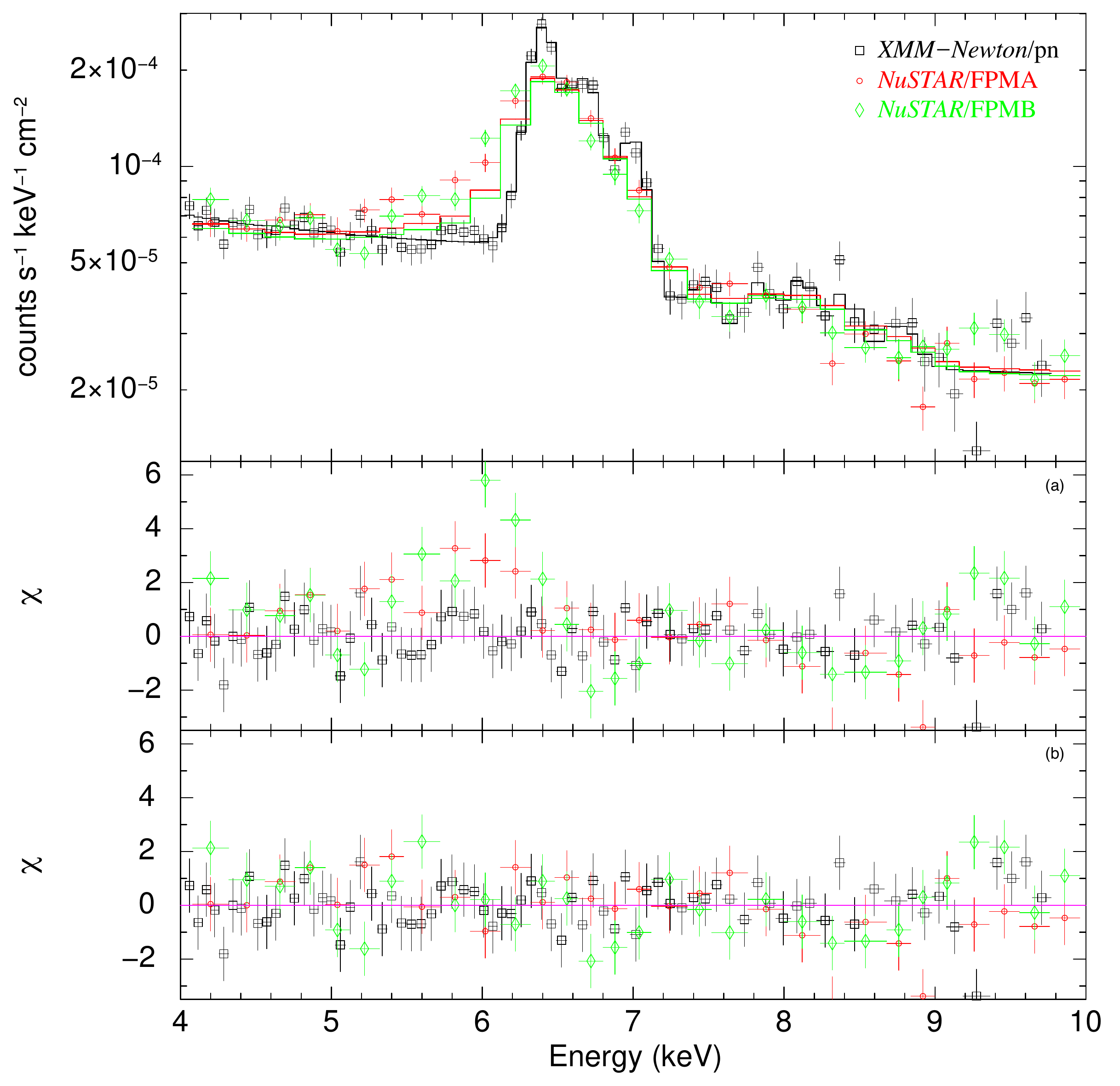}
   \caption{\textit{Upper panel.} \textit{XMM-Newton}/pn (black squares) and \textit{NuSTAR}/FPMA (red circles) and FPMB (green diamonds) spectra of NGC 1068 performed simultaneously in August 2014, showing the 4-10 keV energy range. \textit{Lower panels.} Residuals plotted in terms of sigmas with respect to the phenomenological model (panel (a)) and fitting the \textit{NuSTAR} spectra with an additional emission line (panel (b)). We refer to the text for further details.}
    \label{fig:simult2014}
\end{figure}

As a further step, we take advantage of the simultaneous \textit{XMM-Newton} and \textit{NuSTAR} observations performed in August 2014 (OBSID 0740060401 and 60302003004, respectively), to evaluate the possible presence of a line at $\sim$6 keV in data with higher spectral resolution.
The \textit{XMM-Newton} data were reduced using the latest CCF and the pn spectrum was extracted from a circular region with a 40 arcsec radius centered on the source, and binned in order to oversample the instrumental resolution by at least a factor of 3 and to have no less than 30 counts in each background-subtracted spectral channel.
We note that the extraction region for the source spectrum is smaller than those used to extract the \textit{NuSTAR} spectra, but no sources of emission were present beyond 40" within the EPIC images.
Fitting both \textit{XMM-Newton} and \textit{NuSTAR} observations with a phenomenological model (i.e. \texttt{zpow+pexrav+several zgauss} components in XSPEC), we obtain the best fit in Figure \ref{fig:simult2014}. \textit{XMM-Newton} data are well reproduced by the model, while significant residuals at $\sim$6 keV are clearly visible in \textit{NuSTAR} spectra (see red and green spectral bins in panel (a)).
To reproduce these residuals, an additional \texttt{gauss} component is needed.
It is worth noting that the upper limit to the flux of an emission line at that energy in the \textit{XMM-Newton} spectrum is much lower and inconsistent with the measurement in \textit{NuSTAR} (see Table \ref{tab:line_Aug2014}).

\begin{table}
	\centering
	\small
	\caption{Energy and normalization of the 6 keV line component in the simultaneous \textit{XMM-Newton} and \textit{NuSTAR} observations performed in August 2014. Both spectra are fitted with a phenomenological model.}
	\label{tab:line_Aug2014}
	\begin{tabular}{lcc}
		\hline
							&	E (keV)			&	norm (ph cm$^{-2}$ s$^{-1}$ keV$^{-1}$)	\\
		\hline
		\textit{XMM-Newton}		&	$6.0$$^\dag$		&	$\leq1.9\times10^{-6}$				\\
		\textit{NuSTAR}/FPMA	&	$6.0\pm0.1$		&	$(1.4\pm0.5)\times10^{-5}$			\\
		\textit{NuSTAR}/FPMB	&	$6.1\pm0.1$		&	$(2.5\pm0.6)\times10^{-5}$			\\
		\hline
	\end{tabular} \\
	\raggedright \textit{Note.} $^{\dag}$ Fixed value.
\end{table}

All the previous findings suggest that the significant residuals observed in Model A have no astrophysical origin, but are probably due to calibration issues; therefore, to account for them, we added a spurious emission line (modeled in XSPEC with a \texttt{gauss} component with $\sigma=0$) in our best-fit model (Model E), leaving both its energy and flux free to vary between both the observations and the two \textit{NuSTAR} focal planes.

An analogous feature is also observed in ESO 138-G1 (Zappacosta et al. in prep.), which is another Compton-thick AGN with a very large equivalent width of the Fe K$\alpha$ line (e.g. \citealp{CollingeBrandt2000}, \citealp{Piconcelli2011}, \citealp{DeCicco2015}).

\bsp	
\label{lastpage}
\end{document}